\documentclass[aps,superscriptaddress,reprint,nofootinbib]{revtex4-1}

\usepackage{graphicx}
\usepackage{amsmath}
\usepackage{amssymb}
\usepackage{amsfonts}
\usepackage{filecontents}
\usepackage{color}
\usepackage[normalem]{ulem} 
\usepackage{soul}
\usepackage{epstopdf}
\usepackage{nameref}
\usepackage{fancyhdr}
\usepackage{float}
\usepackage{gensymb}
\usepackage{physics}
\usepackage{upgreek}
\newcounter{fig}
\usepackage{lipsum}

\usepackage[usenames,dvipsnames,svgnames,table]{xcolor}
\usepackage{enumitem}
\usepackage{multirow}

\begin{document}

\title{High Purcell factor generation of indistinguishable on-chip single photons}

\author{F. Liu \normalfont\textsuperscript{$\dag$}}
\affiliation{Department of Physics and Astronomy, University of Sheffield, Sheffield, S3 7RH, United Kingdom}
\author{A. J. Brash \normalfont\textsuperscript{$\dag$}}
\email[Email:]{a.brash@sheffield.ac.uk\\ \normalfont\textsuperscript{$\dag$} F. Liu, A.J. Brash and J. O'Hara contributed equally to this work.}
\affiliation{Department of Physics and Astronomy, University of Sheffield, Sheffield, S3 7RH, United Kingdom}
\author{J. O'Hara \normalfont\textsuperscript{$\dag$}}
\affiliation{Department of Physics and Astronomy, University of Sheffield, Sheffield, S3 7RH, United Kingdom}
\author{L. M. P. P. Martins}
\affiliation{Department of Physics and Astronomy, University of Sheffield, Sheffield, S3 7RH, United Kingdom}
\author{C. L. Phillips}
\affiliation{Department of Physics and Astronomy, University of Sheffield, Sheffield, S3 7RH, United Kingdom}
\author{R. J. Coles}
\affiliation{Department of Physics and Astronomy, University of Sheffield, Sheffield, S3 7RH, United Kingdom}
\author{B. Royall}
\affiliation{Department of Physics and Astronomy, University of Sheffield, Sheffield, S3 7RH, United Kingdom}
\author{E. Clarke}
\affiliation{EPSRC National Epitaxy Facility, Department of Electronic and Electrical Engineering, University of Sheffield, Sheffield S1 3JD, UK}
\author{C. Bentham}
\affiliation{Department of Physics and Astronomy, University of Sheffield, Sheffield, S3 7RH, United Kingdom}
\author{N. Prtljaga}
\affiliation{Department of Physics and Astronomy, University of Sheffield, Sheffield, S3 7RH, United Kingdom}
\author{I. E. Itskevich}
\affiliation{School of Engineering and Computer Science, University of Hull, Hull, HU6 7RX, United Kingdom}
\author{L. R. Wilson}
\affiliation{Department of Physics and Astronomy, University of Sheffield, Sheffield, S3 7RH, United Kingdom}
\author{M. S. Skolnick}
\affiliation{Department of Physics and Astronomy, University of Sheffield, Sheffield, S3 7RH, United Kingdom}
\author{A. M. Fox}
\affiliation{Department of Physics and Astronomy, University of Sheffield, Sheffield, S3 7RH, United Kingdom}

\begin{abstract}

\textbf{\textit{On-chip} single-photon sources are key components for integrated photonic quantum technologies. Semiconductor quantum dots can exhibit near-ideal single-photon emission but this can be significantly degraded in on-chip geometries owing to nearby etched surfaces. A long-proposed solution to improve the indistinguishablility is by using the Purcell effect to reduce the radiative lifetime. However, until now only modest Purcell enhancements have been observed. Here we use pulsed resonant excitation to eliminate slow relaxation paths, revealing a highly Purcell-shortened radiative lifetime (22.7~ps) in a waveguide-coupled quantum dot--photonic crystal cavity system. This leads to  near-lifetime-limited single-photon emission which retains high indistinguishablility ($\textbf{93.9\%}$) on a timescale in which 20 photons may be emitted. Nearly background-free pulsed resonance fluorescence is achieved under $\pi$-pulse excitation, enabling demonstration of an on-chip, on-demand single-photon source with very high potential repetition rates.} 

\end{abstract}

\maketitle

Integrated quantum photonics has made great progress in recent years, with quantum functionality demonstrated in boson sampling and interferometer sensitivity applications~\cite{Aaronson2011}. However, scaling beyond the few-photon level is presently limited by large losses from the use of off-chip single-photon sources (SPSs), with the current state of the art operating at the 3-5 photon level~\cite{Tillmann2013,Broome2013,Wang2017,Loredo2017}. While SPSs have been realized on-chip using four-wave mixing \cite{Silverstone2014}, the very low efficiency imposes significant limitations. A solution to this issue would be to integrate deterministic SPSs on-chip~\cite{PhysRevX.2.011014,PhysRevLett.101.113903,Makhonin2014,Reithmaier2015,Hausmann2012,Sipahigil2016}. Among the possible candidates for such sources, semiconductor quantum dots (QDs) have been shown to offer nearly ideal performance when emitting into free space~\cite{Santori2002,He2013,Somaschi2016,Ding2016a}. In particular, photon indistinguishabilities of $92.1\%$ and $\sim 98\%$ have been demonstrated with microsecond ~\cite{Wang2016b} and nanosecond \cite{Somaschi2016,Ding2016a} photon separation times, respectively.\\

The photon indistinguishability on short timescales is determined by $T_2/(2T_1)$, where $T_1$ is the emitter lifetime and $T_2$ is the coherence time described by $1/T_2=1/(2T_1)+1/T^*_2$. $T^*_2$ is defined as the \textit{pure dephasing} time characterizing the \textit{homogeneous} (Lorentzian) broadening beyond the natural linewidth. The indistinguishability on long timescales can be further reduced by \textit{inhomogeneous} (Gaussian) broadening on a timescale $\gg T_{1}$, e.g. spectral wandering caused by a fluctuating charge environment. The integration of QD sources into on-chip geometries has been observed to significantly reduce photon indistinguishability due to increased charge fluctuations from the nearby etched surfaces~\citep{Makhonin2014,LPOR:LPOR201500321,Liu2017,Kalliakos2016}. A long-proposed~\cite{PhysRevA.69.032305,LPOR:LPOR201500321} approach to overcoming these effects is to use the Purcell effect to enhance the radiative emission rate $ 1/T_1$~\cite{PhysRev.69.674.2,PhysRevLett.95.013904}. 
In theory, strong Purcell enhancement could be obtained by fabricating QDs in cavities with a high $Q$-factor and small mode volume -- such as photonic crystal cavities (PhCCs). However, previously directly measured $T_1$ have only reached $\sim 150$~ps, corresponding to a Purcell factor ($F_\text{P}$) of only $\sim 10$~\cite{doi:10.1063/1.3020295,PhysRevLett.95.013904,PhysRevB.71.241304,Badolato2005a,PhysRevB.66.041303}, over an order of magnitude smaller than the maximum theoretical value. Most studies attribute the large discrepancy to poor spatial overlap between the QD and the cavity mode \cite{Kim2016} or insufficient detector time resolution \cite{PhysRevB.71.241304}. Shorter $T_1\sim 50$~ps indirectly inferred from multiple-parameter fitting was also reported~\cite{Laurent2005,PhysRevB.71.241304}.

\begin{figure*}
\refstepcounter{fig}
	\includegraphics[width=\textwidth]{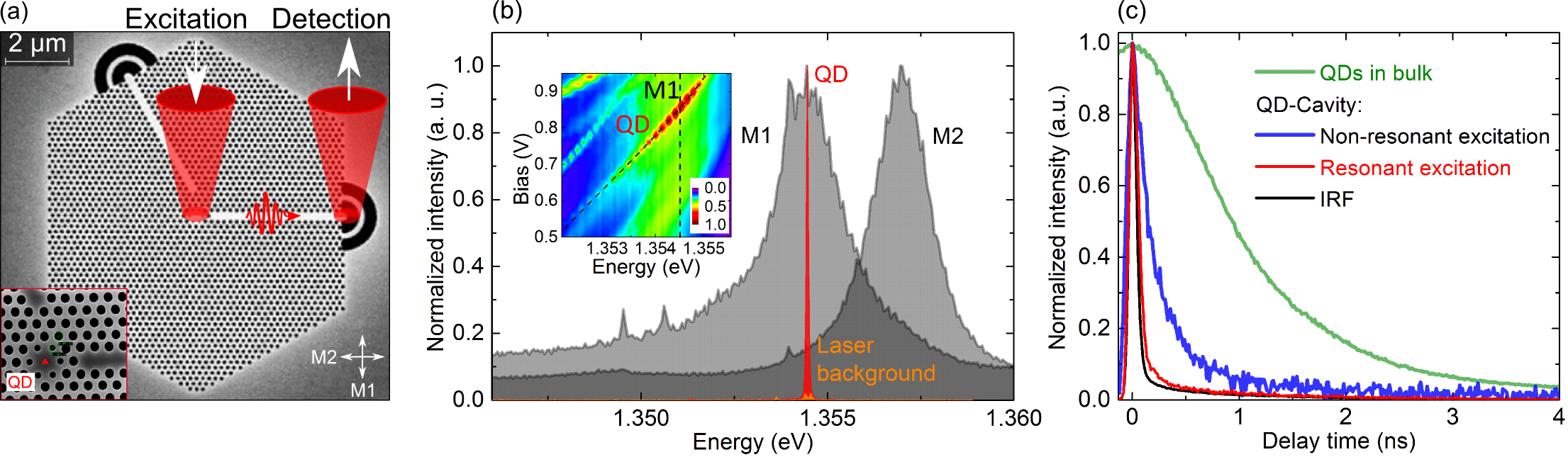}
	\caption{ (a) SEM image of the waveguide-coupled QD--H1 PhCC system. When operated as an on-chip SPS (see Fig.~\ref{HOM}), the QD is excited via the cavity and the single-photon emission is collected from the out-coupler. All other measurements are performed by collecting directly from the cavity to maximize the intensity of the RF signal. Inset: A close-up of the cavity. (b) Grey: High power PL spectra under non-resonant excitation ($ \lambda_\text{exc}=802$~nm). Two orthogonally linearly polarized modes (M1 and M2) are observed when detecting with H and V polarization respectively. Red: Single QD emission measured with resonant $\pi$-pulse excitation. The laser background (orange) is measured by detuning the QD from the laser and is $>20$ times weaker. Inset: Low power PL as a function of the bias and energy under non-resonant excitation. The neutral exciton is electrically tunable by 5.2~meV from bias = 0.2 to 0.93 V (oblique dashed line). Maximum Purcell enhancement of the QD emission is observed around 0.83 V where the QD is resonant with the M1 mode (vertical dashed line). (c) Normalized PL decay of the QD ensemble in bulk measured with non-resonant excitation (green) and that of the QD in cavity measured under non-resonant (blue) and resonant (red) excitation at bias $= 0.83$~V. Black: Instrument response function (FWHM $= 60$~ps).}
	\label{sample}
\end{figure*}

In this article we show unambiguously that larger Purcell enhancements can be achieved by applying pulsed resonant excitation to an InGaAs QD in a waveguide-coupled PhCC. The strongly Purcell-shortened $T_1$ ($ 22.7 \pm 0.9 $~ps) leads to lifetime-limited coherence ($T_2/(2T_1)\approx 1$) and high photon indistinguishability on a timescale in which the source can potentially emit 20 photons. The record-short $T_1$ is directly measured using a new double $\pi$-pulse resonance fluorescence (DPRF) technique and independently verified by resonant Rayleigh scattering (RRS) measurements. Combining very low power $\pi$-pulse excitation and on-chip guiding, we achieve nearly background-free pulsed resonance fluorescence in an on-chip geometry, enabling demonstration of an \textit{on-chip} electrically-tunable SPS meeting three key requirements for quantum information processing: on-demand, high single photon purity ($97.4~\%$) and high indistinguishability ($93.9~\%$). Particularly, the short $T_1$ implies high achievable source repetition rates of $\sim 10$~GHz, crucial for realistic on-chip demultiplexing of the photons.

\section*{Sample design and characterization}

The Purcell factor is determined by the properties of the cavity and the overlap between the QD and the cavity mode, and is given by \cite{PhysRevLett.95.013904}:
\begin{equation}
F_\text{P}=\dfrac{T'_1}{T_1}=\dfrac{3Q}{4\pi ^2 V_m} \dfrac{(2\kappa)^2}{4(\omega - \omega_\text{cav})^2+(2\kappa)^2}\dfrac{\abs{\vec{\mu}\cdot \vec{E}(\vec{r}_0)}^2}{\abs{\vec{\mu}}^2 \abs{\vec{E}_\text{max}}^2}
\label{Eq: FP}
\end{equation}
where $T'_1$ is the exciton radiative lifetime in the absence of a cavity; $Q$ is the quality factor of the cavity, and $V_m$ its mode volume in cubic wavelengths $(\lambda/n)^3$; $\omega$, $\omega_\text{cav}$ and $2\kappa$ denote the angular frequency of the exciton transition, the cavity resonance, and the full width at half maximum (FWHM) of the cavity mode; and $\vec{\mu}$, $\vec{E}(\vec{r}_0)$ and $\vec{E}_\text{max}$ represent the transition dipole moment, the electric field at the QD position and the maximum electric field.\\

In order to obtain strong Purcell enhancement across a large QD tuning range, we integrate the QD into an H1 PhCC with small mode volume ($V_m \approx 0.63 ~(\lambda/n)^{3}$) and moderate $Q$  (see Fig.~\ref{sample}). The cavity has two orthogonally linearly polarized fundamental modes (M1 ($Q = 540$) and M2 ($Q = 765$)) (shaded grey lines in Fig.~\ref{sample}(b)). The upper theoretical limit of the $F_\text{P}$ value is 65 for the M1 mode (see Supplementary Information (SI), section~\ref{dipole}). To extract the photons from the cavity and guide them on-chip, we integrate two W1 photonic crystal waveguides. Each is coupled to one cavity mode \cite{Bentham2015, Coles:14} and terminated with an out-coupler. Integrating the photonic crystal structure into a $p$-$i$-$n$ diode (see SI section~\ref{sec:sample}) allows tuning of the neutral exciton ($X$) by $ \sim 5 $~meV via the quantum-confined Stark effect (see inset in Fig.~\ref{sample}(b)). Clear enhancement of the photoluminescence (PL) intensity is observed when the $X$ is resonant with the M1 cavity mode. 

\begin{figure*}
\refstepcounter{fig}
	\includegraphics[width=0.95\textwidth]{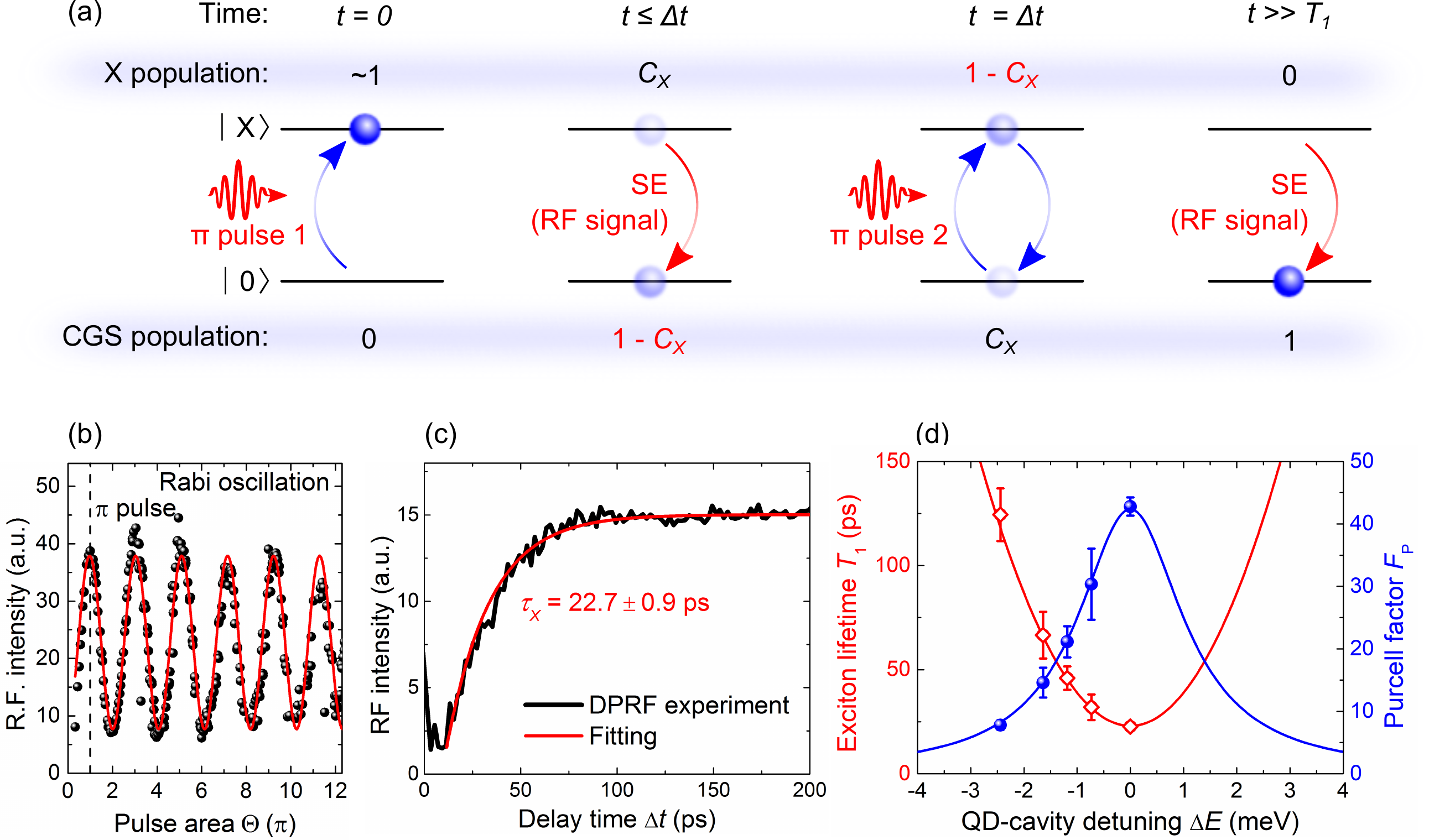}
	\caption{(a) The principle of the DPRF technique. SE: Spontaneous emission. (b) RF intensity of the QD as a function of the pulse area $ \Theta $ of a single pulse, showing Rabi oscillations (red line - sine fit).   $\Theta=\mu/ \hbar \int^{+\infty}_{-\infty}{E(t)\text{d}t}$, where $\mu$ and $ E(t)$ denote the transition dipole moment and the laser field.   (c) DPRF measurement: The RF intensity as a function of the time delay $\Delta t$ between the $ \pi $-pulses. Fitting (red) with a single exponential function gives an exciton radiative lifetime of ($ 22.7 \pm 0.9 $)~ps. (d) The dependence of $T_1$ (red diamonds) and $F_\text{P}$ (blue dots) on the QD--cavity detuning $\Delta E=E_X-E_\text{c}$, where $E_X$ and $E_\text{c}$ are the energies of the exciton and cavity resonances respectively. Solid lines: Simulations using eq.~\ref{Eq: FP}.}
	\label{DPRF}
\end{figure*}

To investigate the Purcell-shortened $T_1$, we first perform time-resolved measurements using a fast single-photon avalanche diode (SPAD).  The PL decay time ($(262 \pm 3)$~ps, blue line in Fig.~\ref{sample}(c)) measured with the QD resonant with the M1 cavity mode under non-resonant excitation ($ \lambda_\text{exc}=802$~nm) is shortened by a factor of $\sim 4$ compared with that of ensemble QDs ($T'_1 = (971 \pm 15)$~ps), a mean value obtained using four different locations outside the photonic crystal (one is shown, green line). The distribution of the QD ensemble peaks at around 1.353~eV, very close to the emission energy (1.354~eV) of the QD on which we focus. Under resonant excitation, the PL decay time of the QD in the cavity is further shortened by at least a factor of 6 (to $(46.2 \pm 1.2)$~ps without deconvolution, red line), a value limited by the instrument response function (IRF) of the SPAD (FWHM~$=60$~ps, black line). We attribute the difference of the PL decay time under resonant and non-resonant excitation to a long carrier relaxation time from higher energy states to the lowest exciton state \cite{doi:10.1063/1.4894239,Zibik2009,Berstermann2007}, supported by simulations (see SI, section~\ref{subsec:decay}). The slow carrier relaxation masks the real $F_\text{P}$ value and limits the indistinguishability of QD SPSs \cite{PhysRevA.69.032305}. This observation implies that in the case of strong Purcell enhancement, $T_1$ can only be accurately measured when the exciton is populated much faster than the radiative recombination rate, in this case by resonant excitation. In addition, since in our sample $T_1$ cannot be clearly resolved by the fastest SPADs available, a technique with higher time-resolution is required.

\section*{Double $\pi$-pulse resonance fluorescence measurement}

To measure $T_1$ accurately, we develop a DPRF technique with a time resolution ultimately limited by the laser pulse duration ($T_\text{P}=13$~ps) (see details in \textit{Methods}, and SI, section~\ref{subsec:DPRF}), making it possible to measure a  $T_1$ much shorter than the time resolution of SPADs. The principle of the DPRF technique is illustrated in Fig.~\ref{DPRF}(a). The QD can be treated as a two-level system consisting of a crystal ground state (CGS) $|0\rangle$ and an exciton state $|X\rangle$ with a total population of 1. At $t=0$, a laser pulse with a pulse area $\Theta=\pi$  coherently drives the QD to $|X\rangle$, creating an $X$ population close to 1. The $\Theta$ is calibrated by performing a Rabi oscillation measurement \cite{Ramsay2010a} (see Fig. \ref{DPRF}(b)). Before the second pulse arrives, the exciton population radiatively decays to $C_X = e^{-\Delta t/T_1}$ via spontaneous emission (SE), where $\Delta t$ is the inter-pulse delay. The probability of photon emission up to time $\Delta t$ is equal to ($1-C_X$). At $t=\Delta t$, the second $\pi$-pulse exchanges the populations of $|0\rangle$ and $|X\rangle$. The exciton population is now ($1-C_X$) which subsequently decays to the ground state.
The total RF intensity ($I_\text{RF}$) measured by the DPRF technique is therefore described by: 

\begin{equation}
I_\text{RF} \propto 2(1-C_X)=2(1-e^{-\Delta t/T_1}).
\label{eq:DPRF}
\end{equation}

Fig.~\ref{DPRF}(c) shows the result of the DPRF measurement at QD--cavity detuning $\Delta E=0$. $I_\text{RF}$ recovers with $\Delta t$ on the timescale of the exciton radiative lifetime. Fitting the curve with eq.~\ref{eq:DPRF} yields a  record-low $T_1$ of $(22.7 \pm 0.9)$~ps, corresponding to a very high Purcell factor for a QD--nanocavity system of $43 \pm 2$ (for $T'_1=971 \pm 15$~ps). The RF signal saturates at a pulse separation of around $100~\textrm{ps}$ in Fig.~\ref{DPRF}(c), opening a route to repetition rates as high as $10~\textrm{GHz}$. Below saturation, there is a significant probability of emitting zero rather than the desired two photons (see SI section~\ref{subsec:DPRF}), defining an upper bound on excitation repetition rate for SPS applications. Unlike for slower sources, on-chip delays of $\sim 100~\textrm{ps}$ can readily be realized \citep{5437515}, paving the way for on-chip time demultiplexing which is an important requirement for integrated photonic circuits. 

Detuning the QD away from the cavity resonance increases (decreases) $T_1$ ($F_\text{P}$) (see Fig.~\ref{DPRF}(d)). This trend is well reproduced by eq.~\ref{Eq: FP} with the cavity linewidth (2.5~meV) extracted from the PL spectra (see Fig.~\ref{sample}(b)) and a spatial overlap of $\sim 81~\%$, further showing that the short $T_1$ results from a large Purcell enhancement.

Our findings demonstrate two advantages of low-$Q$ cavities
for on-chip SPSs. Firstly, although the QD–-cavity coupling strength ($\hbar g$) estimated from the $F_{P}$ value is as large as $135~\upmu \text{eV}$ (see SI, section~\ref{dipole}), the low $Q$ ensures that the system remains in the weak coupling regime, as required for efficient coherent single-photon emission. Secondly, very short $T_{1}$ ($\leq 30 \text{ps}$) may be maintained within a large tuning range (1.4 meV), giving an electrically-tunable source of on-chip single photons.

\section*{Resonant Rayleigh scattering}
\begin{figure}
\refstepcounter{fig}
	\includegraphics[width=\linewidth]{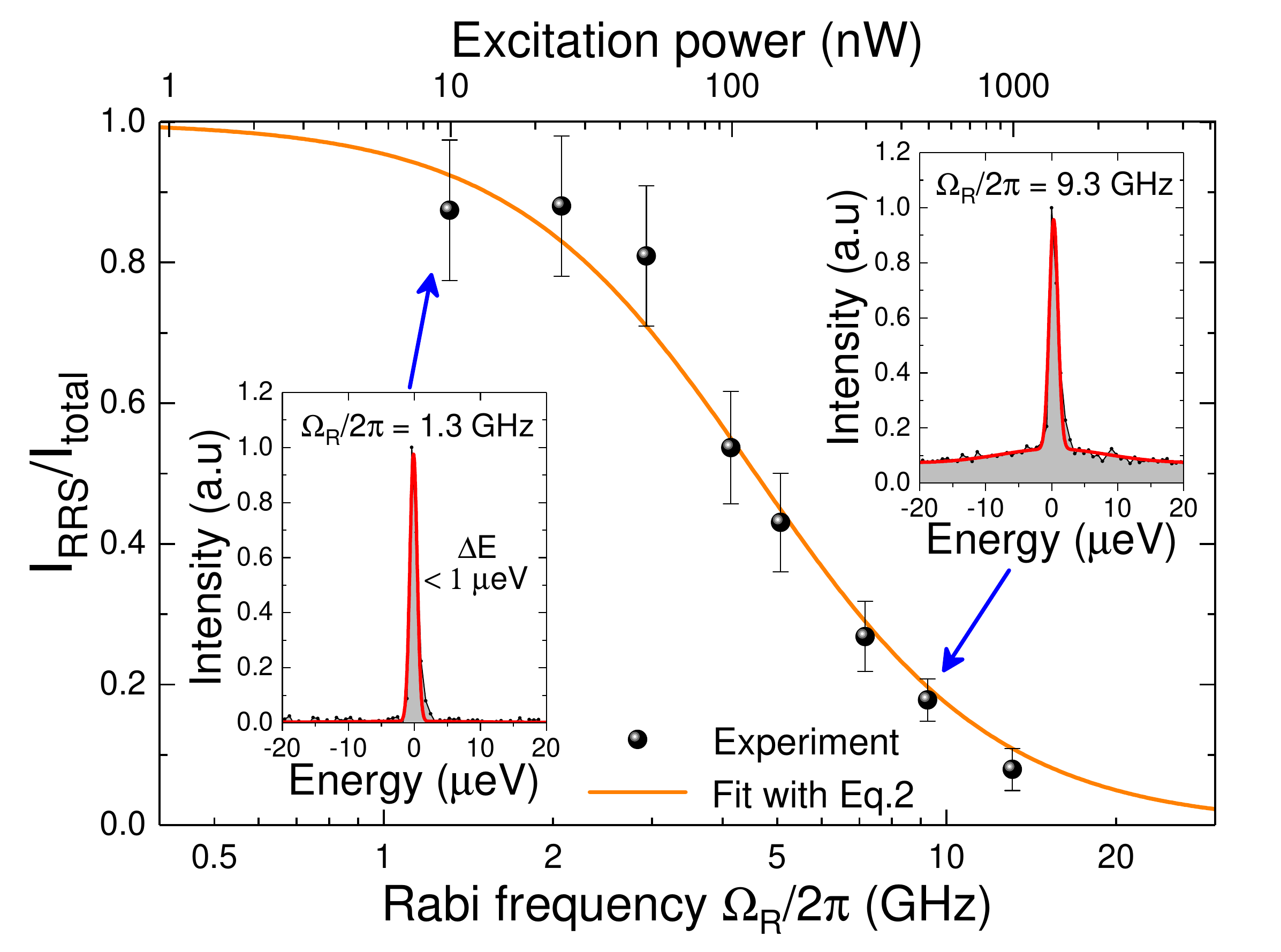}
	\caption{Plot of the ratio of the coherently scattered laser photons ($ I_\text{RRS} $) to the total scatter ($I_\text{total} = I_\text{RRS} + I_\text{SE} $) as a function of Rabi frequency and CW excitation power. Orange line: Fit using eq.~\ref{eq:RRS}. Insets: High resolution spectra of the QD emission under weak (left) and strong (right) CW resonant driving, measured with a Fabry--P\'erot interferometer. Red lines: Fits of the RRS and SE (see SI, section~\ref{subsec:FP}).}
	\label{Rayleigh}
\end{figure}

To further verify the short $T_1$ and probe the pure dephasing of the emitter, we switch to resonant continuous-wave (CW) excitation. The transition is driven at the Rabi frequency $\Omega_\text{R}$, and the exciton population and coherence have damping constants $\gamma_1=1/T_1$ and $\gamma_2=1/T_2$ respectively. In the weak driving limit, where $(\Omega_\text{R})^2 \ll \gamma_1 \gamma_2$, the scattered field is dominated by RRS provided $T_2 > T_1$ \cite{Matthiesen2012,Proux2015,Bennett2016}. These coherently scattered photons are antibunched but retain the linewidth (and thus coherence) of the laser.
The ratio of the RRS intensity to the total (RRS + SE) intensity is given by \cite{Bennett2016}:

\begin{equation}
\dfrac{I_\text{RRS}}{I_\text{total}} = \dfrac{T_2}{2T_1} \dfrac{1}{1 + (\Omega_\text{R})^2/(\gamma_1\gamma_2)}.
\label{eq:RRS}
\end{equation}

Eq.~\ref{eq:RRS} suggests that reducing $T_1$ through a strong Purcell effect will lead to a high fraction of RRS. To demonstrate this, high resolution spectroscopy is performed using a scanning Fabry--P\'erot interferometer (FPI) (see \textit{Methods}). At high driving strengths (Fig.~\ref{Rayleigh}, right-hand inset), the spectrum consists of a sub-$\upmu$eV component from RRS with a broad contribution from SE which vanishes at lower driving strengths (left-hand inset). By fitting the spectra, the ratio $I_\text{RRS}/I_\text{total}$ may be evaluated as a function of $\Omega_\text{R}$.

A fit using eq.~\ref{eq:RRS} (see SI, section~\ref{subsec:FP}) is included in Fig.~\ref{Rayleigh} as an orange line and gives $ T_1 = (24.6 \pm 1.6)~\text{ps}$ and $T_2 = (49.2 \pm 5.4)~\text{ps}$, providing an independent measure of the short radiative lifetime, and showing that the strong Purcell enhancement successfully eliminates the effect of pure dephasing, resulting in close to lifetime-limited coherence ($T_2/(2T_1) \approx 1$).

\section*{On-chip on-demand single-photon source}
\begin{figure}
\refstepcounter{fig}
	\includegraphics[width=\linewidth]{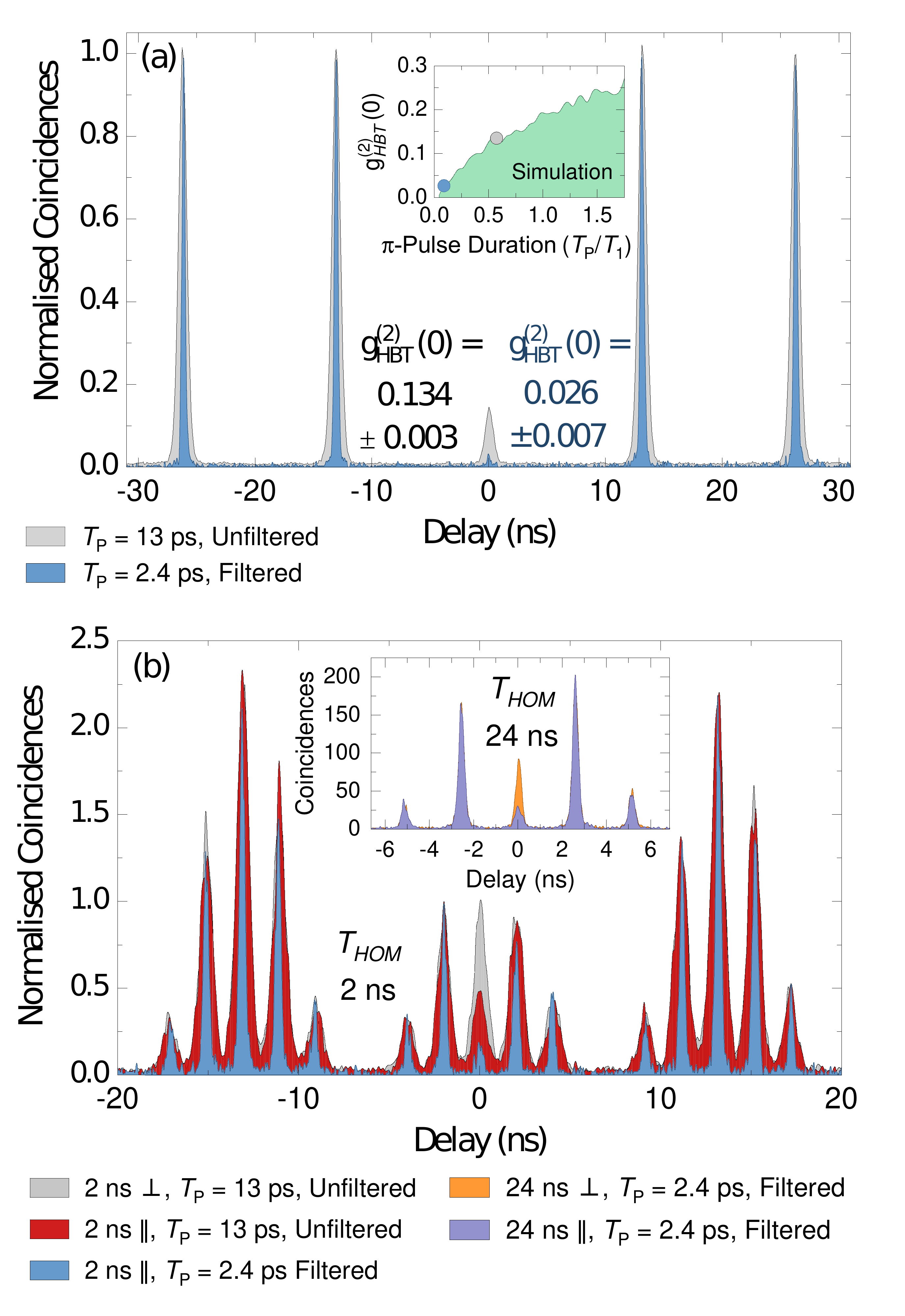}
	\caption{Second-order correlation measurements of the waveguide-coupled QD emission under resonant $\pi$-pulse excitation. (a) Hanbury Brown and Twiss measurement of single-photon purity ($1-\text{g}_{HBT}^{\mathrm{(2)}}(0)$) using 13 ps (grey) or 2.4 ps (blue) pulses. Inset: Simulation of $\text{g}_{HBT}^{\mathrm{(2)}}(0)$ as a function of $\pi$-pulse duration $T_{P}$ relative to $T_{1}$, where coloured circles correspond to experimental data (see SI, section~\ref{subsec:duration-HBT}). (b) Hong-Ou-Mandel measurement of interference visibility for photons emitted $2~\textrm{ns}$ apart. The red and grey data show coincidence counts for co- ($\parallel$) and cross-polarization ($\bot$) of the two interferometer arms respectively when the source is driven by 13 ps pulses. The blue data shows the co-polarized case for 2.4 ps pulses. Inset: HOM measurement for photons emitted $24~\textrm{ns}$ apart. Note that the peak area pattern differs from the main figure as $(1/T_{HOM}) < 76.2~\text{MHz}$ (the laser repetition rate), thus the peaks adjacent to zero delay now originate from different pulses. See \textit{Methods} and SI, section~\ref{sec:correlation} for further details.}
	\label{HOM}
\end{figure}

In order to generate strings of single photons on-demand we now study our device under resonant $\pi$-pulse resonant excitation. QDs driven by $\pi$-pulses have proven to be an excellent source of single photons owing to their high purity, indistinguishability and on-demand operation \cite{He2013,Somaschi2016,Ding2016a}. Such performance would be highly desirable for an on-chip SPS. However, to date all QD SPSs driven by resonant $\pi$-pulses have emitted into free-space. By exciting on the cavity and collecting from the waveguide (see Fig. \ref{sample}(a)), we achieve nearly background-free pulsed RF (see red and orange lines in Fig.~\ref{sample}(b)), realizing a resonantly-driven on-chip on-demand SPS. Compared with QDs in bulk or relatively large nanostructures, it is significantly more experimentally demanding to realize background-free pulsed RF in photonic crystal structures because the patterned surface scatters the polarization of the reflected laser.

To characterize the purity of the source, a Hanbury Brown and Twiss (HBT) correlation measurement is performed under resonant $\pi$-pulse excitation. The results are shown in Fig. \ref{HOM}(a) where the area of the grey time-zero peak for a 13 ps pulse gives a purity ($1-\text{g}_{HBT}^{\mathrm{(2)}}(0)$) of $86.6 \pm 0.3~\%$ at an unfiltered signal-to-background ratio (SBR) $\approx$~20:1. Simulations (inset to Fig. \ref{HOM}(a), see also SI, section~\ref{subsec:duration-HBT}) show that the measured single-photon purity is limited primarily by multiple emissions originating from re-excitation of the source by a pulse that is relatively long compared to $T_{1}$. To test this hypothesis and suppress multiple emissions during the pulse, the measurement is repeated with a 2.4 ps pulse (blue data in Fig. \ref{HOM}(a)). Owing to intrinsic birefringence of the optical setup, a $96~\upmu\text{eV}$ grating filter is required to eliminate residual scatter of the spectrally broad pulse from the sample surface, resulting in an SBR $\approx$~50:1. We emphasize that such filtering is only required because of the combination of out-of-plane collection geometry and relatively short ($5~\upmu$m) waveguide length; this would not be required for on-chip experiments. In agreement with simulations, the measured purity increases to $97.4 \pm~0.7~\%$ with the shorter pulse. For 13 ps pulses, the filtered and unfiltered purities are very similar, indicating that the purity is improved by the reduced pulse duration rather than the filtering.

Using a fiber Mach-Zehnder interferometer (see SI, section~\ref{sec:correlation}), Hong-Ou-Mandel (HOM) interferometry is performed to determine the indistinguishability of photons emitted from the source (Fig. \ref{HOM} (b)). When the photon separation ($T_{HOM}$) is 2 ns and a 13 ps pulse is used without filtering, the visibility ($V$) is $(60.1\pm 3.2)~\%$ after correcting for the interferometer properties (see SI, section~\ref{sec:correlation}). If $\text{g}_{HBT}^{\mathrm{(2)}}(0)$ is also corrected for, this rises to $V = (79.7 \pm 5.9)~\%$. By again reducing the pulse duration to 2.4 ps, the visibility increases to $(89.4\pm 2.5)~\%$ ($(93.9\pm 3.3)~\%$) without (with) correction for $\text{g}_{HBT}^{\mathrm{(2)}}(0)$, implying a $T_{2}/(2T_{1})$ ratio close to unity, in agreement with the RRS measurements.

The improved visibility with the 2.4 ps pulse is mainly due to the previously discussed reduction of multiple emission events, although the spectral filter also acts to remove a significant amount of the phonon sideband. Recent studies have indicated that the unfiltered visibility of single photons from non-Purcell-enhanced InGaAs QDs at $4.2 ~\text{K}$ is limited to around $80~\%$ by incoherent phonon sideband emission \cite{2016arXiv161204173I}. This can be improved without the losses of filtering by placing the QD in a resonant high-$Q$ cavity \cite{2016arXiv161204173I}. In the device studied here, whilst there is a strong Purcell enhancement, the relatively low $Q$ means that the cavity filtering effect is weaker, introducing a theoretical upper bound on the unfiltered visibility of $\sim 90~\%$, rising to $\sim 99~\%$ if the grating filter is added \cite{2016arXiv161204173I}. A separation of $T_{HOM}=2~\text{ns}$ would correspond to 20 emission cycles of the source (if driven at $10~\text{GHz}$), adequate to significantly exceed the complexity of any boson sampling experiments to date ~\cite{Tillmann2013,Broome2013,Wang2017,Loredo2017}.

To explore any potential degradation of the visibility at longer timescales, the separation is extended to $T_{HOM}=24~\text{ns}$ (potentially 240 emission cycles) (see inset to Fig. \ref{HOM}(b)). This results in a visibility of $(75.3\pm 2.0)$~\% ($(79.9\pm 3.4)$~\%) without (with) correction for $\text{g}_{HBT}^{\mathrm{(2)}}(0)$, a decrease of 14~\% compared to $T_{HOM}=2~\text{ns}$. As this timescale is $\gg T_{1}$, the decline in visibility is attributed to spectral wandering due to a charge environment fluctuating on the timescale of tens of nanoseconds. Previous studies of QD microlens structures (which also include etched surfaces relatively close to the QD) exhibited a significantly larger wandering-induced visibility decay of $\sim 40~\%$ on a comparable timescale ($12.5~\textrm{ns}$) \citep{PhysRevLett.116.033601}. The critical advantage of the device studied here is that the very short $T_{1}$ broadens the natural linewidth by a factor of $F_{P}$, minimising the visibility degradation 
whilst also allowing photons to be extracted much faster than spectral wandering timescales.

\section*{Discussion}

For on-chip single-photon sources, reduced photon indistinguishability through environmental interaction has been a major concern. This is especially true for waveguide-coupled sources, which by necessity are situated near surfaces \cite{wang_optical_2004}. In this paper, the effect of pure dephasing on the waveguide-coupled QD emission has been made negligible through use of the Purcell effect and resonant excitation, as is shown by the high RRS fraction and high HOM visibility for short pulse separations.

Another potential issue, as the comparison of different HOM photon separations indicates \cite{PhysRevLett.116.033601,Wang2016b,Loredo:16}, is wandering due to a fluctuating charge environment. This is also mitigated by the Purcell enhancement, since the ratio between the lifetime-limited linewidth of the QD emission and the width of the wandering is reduced by a factor of $F_\text{P} \sim 40$. We note that this is a first-generation device, and further improvement of the indistinguishability at long photon separation times could potentially be achieved by reducing the charge fluctuations via surface passivation~\cite{Liu2017} or by optimizing the sample growth and diode structure~\cite{PhysRevB.96.165440}. We also note that keeping all other parameters constant, increasing $Q$ to 2500 (the onset of strong coupling) by optimizing fabrication would give $F_{P}\sim 200$ (see SI section~\ref{dipole}), further suppressing the influence of spectral wandering, while also improving the theoretical unfiltered visibility to $\sim 97\%$ by reducing the phonon sideband content of the emission \cite{2016arXiv161204173I}. In off-chip experiments driven by QD SPSs, visibilities of $\sim 65~\%$ have been sufficient to demonstrate boson sampling \citep{Loredo2017,Loredo:16}, with $94~\%$ being the current state of the art \citep{Wang2017}. This confirms the feasibility of harnessing our source architecture to perform such quantum optics experiments on a single chip.

Besides indistinguishability, the count rate measured by a detector is another important figure-of-merit. Using experimentally demonstrated parameters for the GaAs platform (see SI section \ref{sec:brightness}), the count rate is predicted to be $\sim 4$~MHz for a SPS driven at $76.2~\text{MHz}$ and connected to a superconducting nanowire single photon detector (SNSPD) via a $100~\upmu$m photonic crystal waveguide. This is comparable with the highest count rate (9~MHz) of micropillar-based off-chip SPSs~\cite{Wang2017}. Thanks to the large Purcell enhancement, the maximum count rate for our source can potentially reach $\sim 540~\text{MHz}$ when driven with a pulse repetition rate of 10~GHz. Beyond this, optimizing the cavity--waveguide coupling \cite{Coles:14}, improving the SNSPD efficiency \cite{Pernice2012} and increasing the cavity $Q$ presents a clear path to GHz on-chip count rates, showing the great potential of this approach for integrated quantum photonics.

\section*{Conclusion}

In this article we unambiguously reveal a strongly Purcell-shortened exciton radiative lifetime of only 22.7~ps in a photonic crystal cavity using pulsed resonant excitation. This is directly measured by a novel high-time-resolution DPRF technique. Electrically tunable on-demand single photons from the cavity are efficiently channeled into a waveguide with minimal laser background, allowing the device to operate as an on-chip SPS. The short radiative lifetime ($T_1$) opens the way to source repetition rates $\sim 10~\textrm{GHz}$ which are compatible with on-chip delays for time demultiplexing \citep{5437515} and could lead to detected on-chip count-rates of $\sim 540~\text{MHz}$ using experimentally demonstrated parameters.

Additionally, the small $T_1$ eliminates the effect of pure dephasing and suppresses the influence of spectral wandering. This leads to lifetime-limited emitter coherence and high single-photon purity (97.4~\%). Indistinguishabilities of $>90~\%$ are measured on a timescale of 2 ns (potentially 20 photon emission events when driven at $10~\text{GHz}$) or $\sim 80~\%$ for 24 ns (240 photons), sufficient for a future single-chip device to perform fully-integrated quantum optics experiments such as boson sampling~\cite{Loredo2017,Wang2017} with high photon numbers.  Other important QIP proposals such as fast single-photon switching \cite{PhysRevA.82.063821} and photonic cluster state generation~\cite{PhysRevLett.103.113602} will also benefit significantly from a short $T_1$.

Our work demonstrates that a high-performance QD-based SPS can be realised in a scalable on-chip geometry, requiring orders of magnitude less excitation power and space than existing spontaneous four-wave mixing sources~\cite{Silverstone2014} and benefiting from on-demand operation and a much higher photon generation rate. As such, our on-chip source has the potential to be a major step forward in fully-integrated chip devices for quantum photonics~\cite{LPOR:LPOR201500321}.

\section*{Methods} 

\label{sec:methods}

\subsection*{DPRF Setup}

The QD is resonantly driven by a pair of variable duration pulses derived by splitting and Fourier transform shaping a broad 100~fs laser pulse generated from a Ti:Sapphire laser with repetition rate 76.2~MHz. The Gaussian pulse width may be varied by adjusting the width of a slit placed slightly defocused from the Fourier plane. For most experiments, a duration of 13 ps is chosen to maximise the unfiltered signal-to-background ratio (by reduced spectral width) whilst remaining shorter than the QD radiative lifetime.

\begin{figure}[h]
\includegraphics[width=0.45\textwidth]{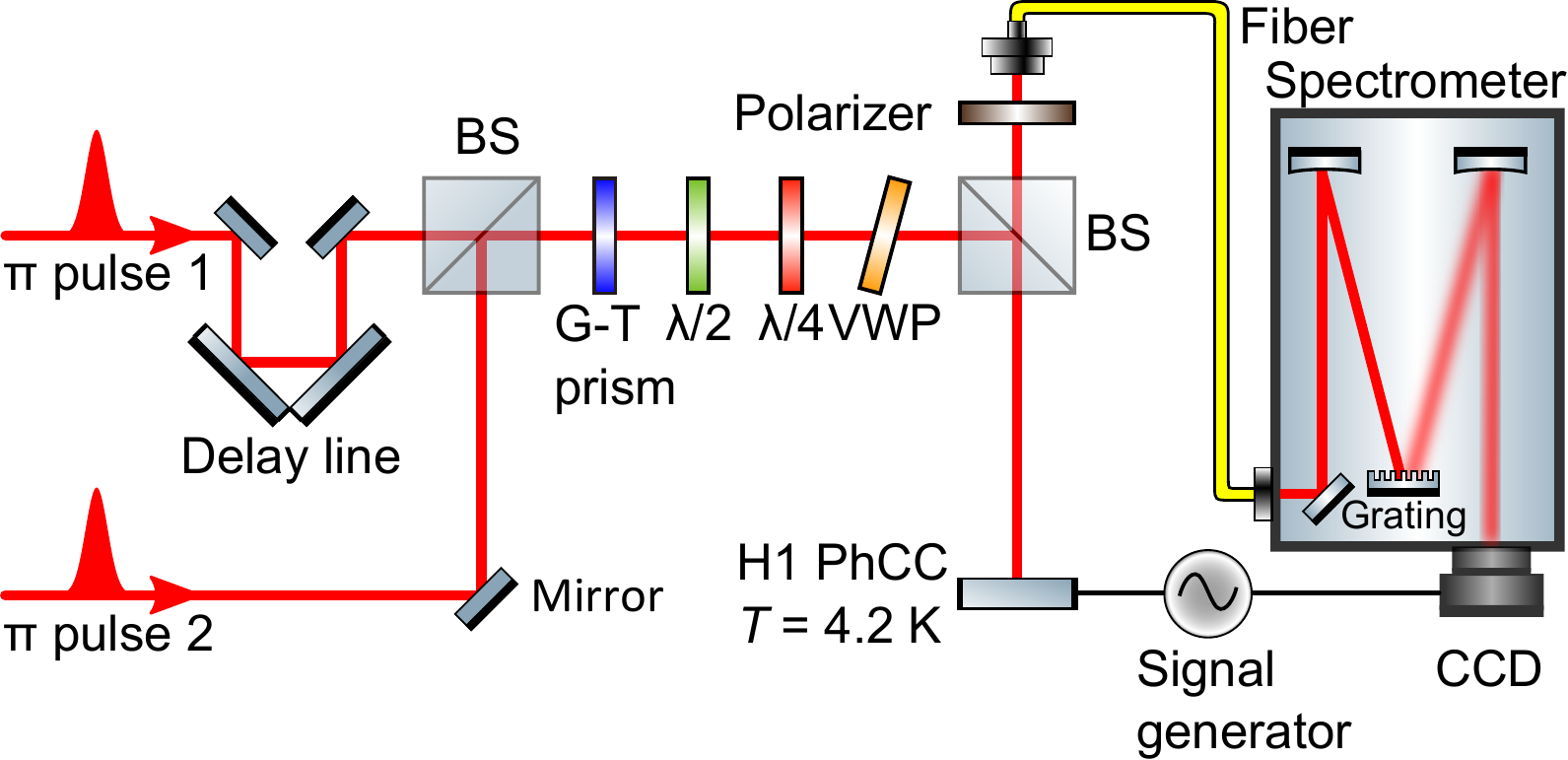}
\caption{The optical setup used for the experiments.}
\label{fig:Setup}
\end{figure}

A cross-polarization configuration is adopted to detect the resonant QD emission, as shown in Fig.~\ref{fig:Setup}. The polarization direction of the laser pulses is initially defined by a Glan--Taylor prism, rotated by a $\lambda/2$ plate and reflected by a non-polarizing beam splitter (BS). The combination of the $\lambda/2$ plate and the BS allows us to easily set the polarization of the laser pulse. For these measurements, the laser pulses are $45\degree$ polarized with respect to the M1 cavity mode. The reflected laser is filtered out by a cross-polarizer. The distortion of the polarization of the laser by all optical components in the excitation \textit{and} detection paths is corrected by a $\lambda/4$ plate and an additional tunable wave-plate with quarter-wave phase retardation (VWP).

The spectrally-integrated signal to background ratio under $\pi$-pulse excitation is $\sim$20:1, smaller than that ($\sim$150:1) under CW excitation (laser power $=25$~nW) due to difficulties in rejecting a broadband laser pulse using polarization. To fully separate the RF signal from the laser background in the DPRF measurement, the bias of the diode is modulated with a frequency of 11~Hz to move the QD in and out of resonance with the laser pulse. The laser background can be fully removed by subtracting the two spectra  from each other (see example QD and background spectra in Fig.~\ref{sample}(b)).

\subsection*{SPAD Lifetime Measurements}

The single-photon avalanche diode (SPAD) lifetime measurements are performed using the optical setup of Fig. \ref{fig:Setup} but using only a single excitation pulse. For the ensemble lifetime of QDs outside the photonic crystal, the excitation is provided by the unshaped ($\sim 100~\textrm{fs}$) output of the Ti:S laser operating at $\lambda = 780~\textrm{nm}$. A $900~\textrm{nm}$ long-pass filter is inserted after the detection polarizer to remove the laser and any wetting layer emission from the detection path. The collection fiber is connected directly to a SPAD operating in Geiger mode with a Gaussian IRF of FWHM $350~\textrm{ps}$. A time-correlated single-photon counting module (TCSPCM) synchronized with the laser pulse train records the arrival times of individual photons to produce the decay curves. For the QD-cavity lifetime measurements the zero-phonon line is filtered through the spectrometer ($94~\upmu \textrm{eV}$ bandwidth) before passing to a different SPAD with higher time resolution (IRF $\sim 60~\textrm{ps}$ with a weak, longer tail) and being analyzed by the TCSPCM as before. For the above-band lifetime measurement the excitation pulse is supplied by the unshaped laser at $\lambda = 802~\textrm{nm}$ whilst the resonant $\pi$-pulse is provided by a single pulse-shaper as in the DPRF measurement but with the second pulse blocked.

\subsection*{Resonant Rayleigh Scattering}

For the RRS measurements a narrow-linewidth ($< 50~\textrm{kHz}$) continuous-wave tunable Ti:S laser provides the excitation source. After the laser, the optical setup is as in Fig. \ref{fig:Setup} except that the emission is passed to the exit slit of the spectrometer and filtered as previously described. The emission then passes through a scanning Fabry--P\'erot interferometer (FPI) and is detected with a SPAD. The FPI is swept by a function generator which also provides a synchronization signal to the TCSPCM, allowing conversion from SPAD detection time to spectral position. The excitation power is converted to $\Omega_\text{R}$ by measuring the power-dependent splitting of the Mollow triplet (see SI, section~\ref{MtaRf}).

\subsection*{Correlation Measurements}

To perform the correlation measurements, the optical setup described in Fig. \ref{fig:Setup} is used. For measurements with the 13 ps pulse, the detection fiber is connected directly (bypassing the spectrometer) to a fiber Mach--Zehnder interferometer. One arm of the interferometer incorporates a $\lambda/2$ wave-plate and the other an additional length of fiber corresponding to a delay of $T_{HOM}$. Further details of the interferometer are contained within the SI, section~\ref{sec:correlation}. The two output ports of the interferometer are connected to a pair of single-photon avalanche photodiodes (combined Gaussian IRF has FWHM $860~\mathrm{ps}$), which in turn are fed to the TCSPCM in order to measure the number of coincidence counts. For the 2.4 ps pulse, the spectrometer provides the additional filtering of the emission ($96~\upmu \textrm{eV}$ FWHM with Gaussian profile) and a pair of single-photon avalanche photodiodes with faster timing response are used (combined Gaussian IRF with $341~\mathrm{ps}$ FWHM).

For HBT measurements, a single $\pi$-pulse per laser cycle ($13.2~\mathrm{ns}$) is applied to the sample (the second pulse is blocked) and only the second fiber splitter of the interferometer is used. For HOM measurements the full interferometer is used and a pair of $\pi$-pulses is applied to the sample as in the DPRF experiment. The pulse separation is matched to the interferometer delay by connecting the two pulses directly to the interferometer, scanning the delay line and observing the maxima of the classical interference between the two pulses. 

\section*{Acknowledgement}
This work was funded by the EPSRC (UK) Programme Grants EP/J007544/1 and EP/N031776/1. The authors thank A. Ul-Haq, J. Iles-Smith, G. Buonaiuto, R. Kirkwood, and S. Hughes for helpful discussions.  

\section*{Author contributions}
F.L. and A.J.B. designed and oversaw the experimental program. A.J.B., L.M.P.P.M. and F.L. developed the DPRF technique and carried out the measurements. J.O'H., L.M.P.P.M., A.J.B. and F.L. performed the SPAD lifetime measurements. J.O'H. and A.J.B. performed the RRS measurements with additional input from N.P.. A.J.B., J.O'H., L.M.P.P.M., F.L. and C.L.P. performed the pulsed correlation measurements. J.O'H. performed the master equation simulations of the system. R.J.C. designed the photonic structures and performed FDTD simulations of them. C.B. and I.E.I. performed initial characterisation of the sample. E.C. grew the quantum dot wafer whilst B.R. fabricated the photonic nanostructures and processed the QD wafer into diodes with assistance from C.B.. L.R.W, I.E.I., M.S.S and A.M.F. provided supervision and expertise. F.L., A.J.B., J.O'H. and A.M.F. wrote the manuscript with input from all authors.

\section*{Additional information}
Supplementary infomation is available in the online version of the paper. Reprints and permission information is available online at URL. Correspondence and requests for materials should be addressed to a.brash@sheffield.ac.uk. F. Liu's present address: JARA-Institute for Quantum Information, RWTH Aachen University, D-52074 Aachen, Germany; N. Prtljaga's present address: Gooch $\&$ Housego (Torquay), Broomhill Way, Torquay, TQ2 7QL, United Kingdom.

\section*{Competing Interests}
The authors declare that they have no competing financial interests.

\bibliography{References}
\bibliographystyle{naturemag}

\setcounter{equation}{0}
\setcounter{figure}{0}
\setcounter{table}{0}

\renewcommand{\theequation}{S\arabic{equation}}
\renewcommand{\thefigure}{S\arabic{figure}}%
\renewcommand{\thetable}{S\arabic{table}}%

\onecolumngrid
\newpage

\begin{Large}
\begin{center}
\textbf{Supplementary Materials: High Purcell factor generation of coherent on-chip single photons}
\end{center}
\end{Large}

\pagenumbering{gobble}

\section{Sample structure}
\label{sec:sample}

\begin{figure}[h]
\includegraphics[width=0.4\textwidth]{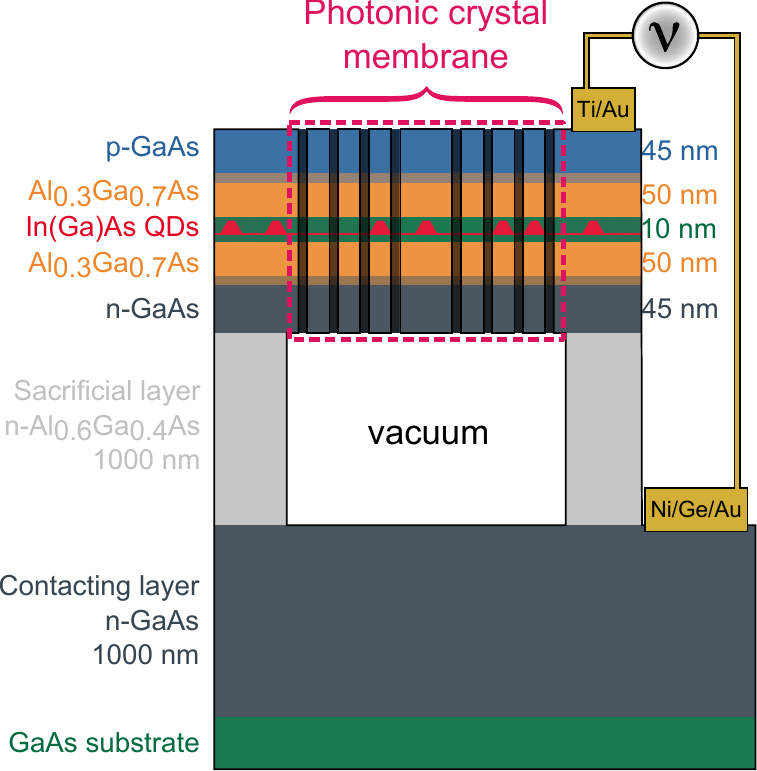}
\caption{Sample structure. A photonic crystal membrane fabricated on a $p$-$i$-$n$ diode structure containing InGaAs QDs.}
\label{fig:Structure}
\end{figure}

Fig.~\ref{fig:Structure} shows the cross-sectional view of the sample. The photonic crystal structure is integrated into a $p$-$i$-$n$ diode allowing tuning the exciton via the quantum-confined Stark effect.

\section{QD--waveguide coupling efficiency}
\label{sec:efficiency}

The coupling efficiency between the M1 cavity mode and the waveguides $\eta_\text{c-u}$ can be estimated according to $\eta_\text{c-u} = 1-Q_\text{M1}/Q_\text{u}$ (Coles et al., Optics Express, 22, 3, 2014), where $Q_\text{M1}$ (540) denotes the $Q$ factor of the M1 mode; $Q_\text{u}$ (1109) is the measured average $Q$ factor of cavities fabricated without waveguides on the same sample. The total coupling efficiency between the M1 mode and the two waveguides is therefore $51\%$.

The coupling efficiency from the M1 mode to each waveguide (see Fig. 1(b)) is $41\%$ and $10\%$ respectively, estimated from the ratio (4:1) of the QD PL intensity measured from the two out-couplers when the QD is resonant with M1. FDTD simulations (Coles et al., Optics Express, 22, 3 2014) show that a maximum theoretical coupling efficiency of up to $89~\%$ between the cavity mode and the waveguide could be achieved in an optimized device.

Finally, the QD--waveguide coupling efficiency $\eta_\text{q-u}$ can be estimated for this device according to $\eta_\text{q-u}=\beta \times \eta_\text{c-u}=40\%$, where $\beta=F_\text{P}/(1+F_\text{P})=98\%$ is the QD-cavity coupling efficiency and $\eta_\text{c-u}=41\%$ is the cavity-waveguide coupling efficiency.

\section{Exciton fine-structure splitting and eigenstate orientation}
\label{sec:FSS}

\begin{figure}[h]
\includegraphics[width=0.6\textwidth]{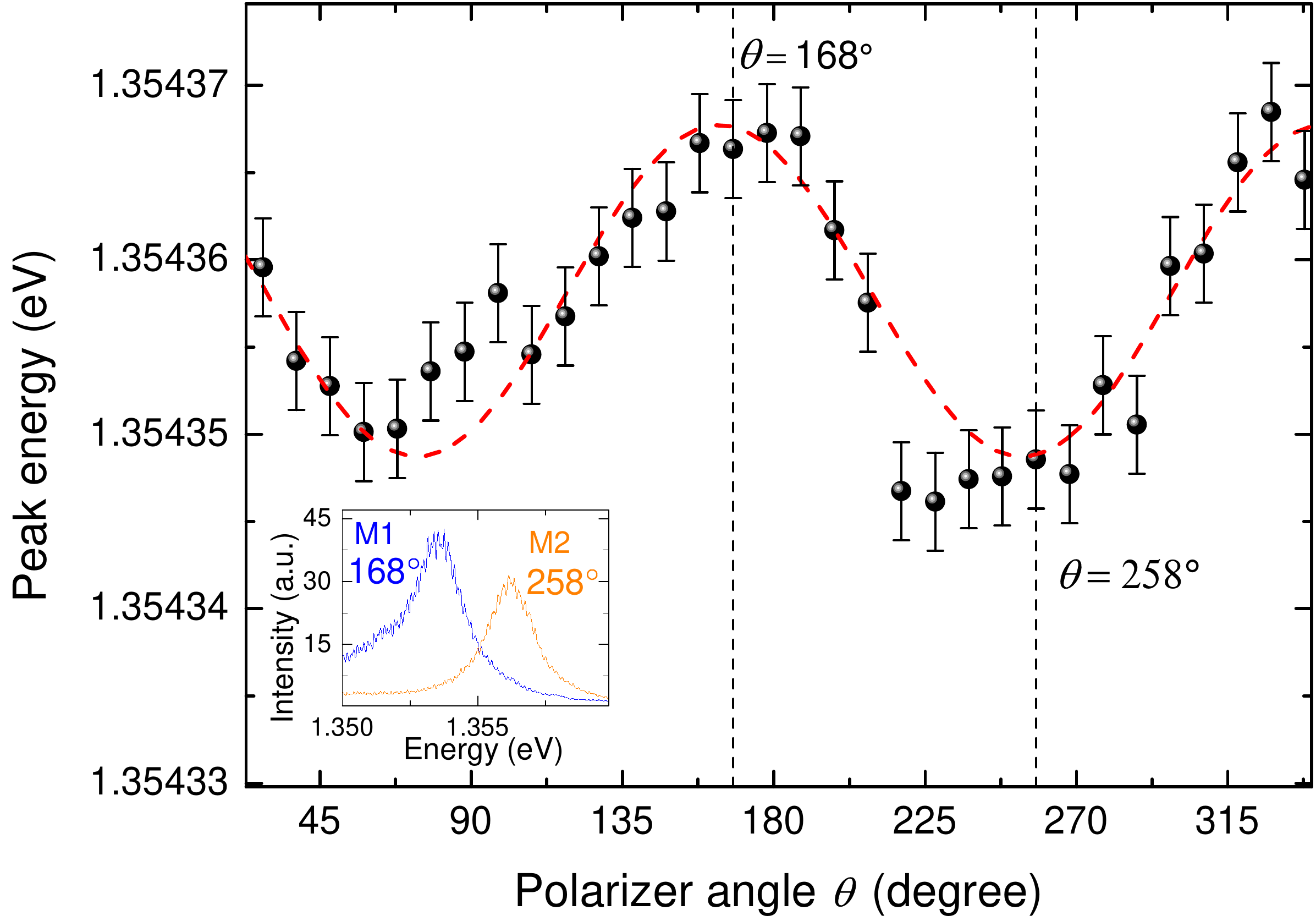}
\caption{Peak energy of the QD emission versus the angle ($\theta$) of the detected polarization. Red dashed lines: guide for the eye. Inset: PL spectra of the two cavity modes measured when the polarizer is co-polarized with mode M1 (blue line), and mode M2 (orange line).}
\label{fig:FSS}
\end{figure}

The charge species of the studied exciton is identified by measuring the exciton fine-structure splitting (FSS). Fig.~\ref{fig:FSS} shows the peak energy of the QD emission as a function of the angle ($\theta$) of the collection polarization. A FSS of $19~\upmu\text{eV}$ is clearly observed, illustrating that the exciton under study is a neutral exciton. 

The inset shows high power PL spectra of the two cavity modes measured when the polarizer is co-polarized with the M1 (blue line, $\theta=168^\circ$) and M2 (orange line, $\theta=258^\circ$). Note that the two QD eigenstates are co-polarized with the two cavity modes respectively, which is expected since both the QD eigenstates and the fundamental modes of the H1 PhCC were intended to be aligned parallel/perpendicular to the (110) crystal axes.

\section{Dipole coupling strength, position and orientation}
\label{dipole}

At zero QD--cavity detuning and for perfect dipole positioning and orientation, the Purcell factor is
\begin{equation}
F_\text{P}=\dfrac{3}{4\pi ^2}\dfrac{Q}{V_m}=\dfrac{2g^2}{\kappa\gamma'_1}=2C,
\label{Eq: FP2}
\end{equation}
where $Q$ is the quality factor, $V_m$ the mode volume in cubic wavelengths $(\lambda/n)^3$, $\hbar g$ the QD--cavity coupling strength ($\upmu$eV), $2\hbar\kappa$ the cavity linewidth ($\upmu$eV), $\hbar\gamma'_1$ the QD's natural linewidth ($\upmu$eV), and $C$ the cooperativity. $Q$ (and $\kappa$) are known from a high-power PL measurement, and $V_m$ is taken from FDTD simulations approximating the real fabricated system rather than the ideal H1 value (giving $0.63$ rather than $0.39~(\lambda/n)^3$). These $Q$ and $V_m$ values give the ideal $F_\text{P}$ for the fabricated cavity as $65$. Then, using the ensemble lifetime $T'_1$ of QDs outside the cavity to obtain $\gamma'_1=1/T'_1$, $\hbar g$ is calculated to be $166~\upmu$eV for the ideal $F_\text{P}$ (i.e. for ideal coupling), and $135~\upmu$eV for the measured QD--cavity system with $F_\text{P}=43$, through (Khitrova et al., Nat. Phys. 2, 81-90, 2006):

\begin{equation}
g=\sqrt{\dfrac{\omega\left|\vec{\epsilon}(\vec{r}_0)\cdot\vec{\mu}\right|^2}{2\hbar \varepsilon_0 n^2 V_m}},
\label{Eq: g}
\end{equation}
and
\begin{equation}
|\vec{\mu}|=\sqrt{3\pi\hbar\varepsilon_0\dfrac{\gamma'_1 c^3}{n \omega^3}}.
\label{Eq: dm}
\end{equation}
The calculated QD dipole moment from eq. \ref{Eq: dm} is $|\vec{\mu}|=27.2~\text{D}$. $\vec{\epsilon}(\vec{r}_0)$ is the field at the QD position normalized to the cavity field maximum $\vec{E}(\vec{r}_0) / \vec{E}_\text{max}$. Then, knowing that for the measured Purcell factor we have $\hbar g=135$~$\upmu$eV, where the maximum is $166~\upmu$eV, it follows that $\left|\vec{\epsilon}(\vec{r}_0)\cdot\vec{\mu}\right|^2/|\vec{\mu}|^2 = 0.81^2$, i.e. the spatial overlap and alignment of the QD dipole and the cavity mode is $\sim 81$~\% ideal. The high coupling is shown by both the very short lifetime and the very large Mollow splitting, discussed in section~\ref{MtaRf}. The large cavity loss does however prevent the system entering the strong-coupling regime, i.e. vacuum Rabi-splitting. This occurs when (Reithmaier et al., Nature 432, 197-200, 2004):

\begin{equation}
16g^2 > (2\kappa-\gamma'_1)^2,
\label{Eq: VRS}
\end{equation}
a condition not satisfied for this $g$ and $\gamma'_1$ until $Q>2500$ and $F_\text{P}\sim 200$. The system thus remains in the weak coupling regime despite the large coupling strength. In general we want $\kappa/2 > g\gg \gamma'_1$ in order to obtain a highly coherent on-chip single-photon source. The device we report here has $\hbar\{2\kappa,g,\gamma'_1\} = \{2510, 135, 0.68\}~\upmu$eV.

\pagenumbering{arabic}
\setcounter{page}{2}


\section{Influence of the cavity on excitation efficiency}
\label{sec:excitation}

\begin{figure}[h]
\includegraphics[width=0.6\textwidth]{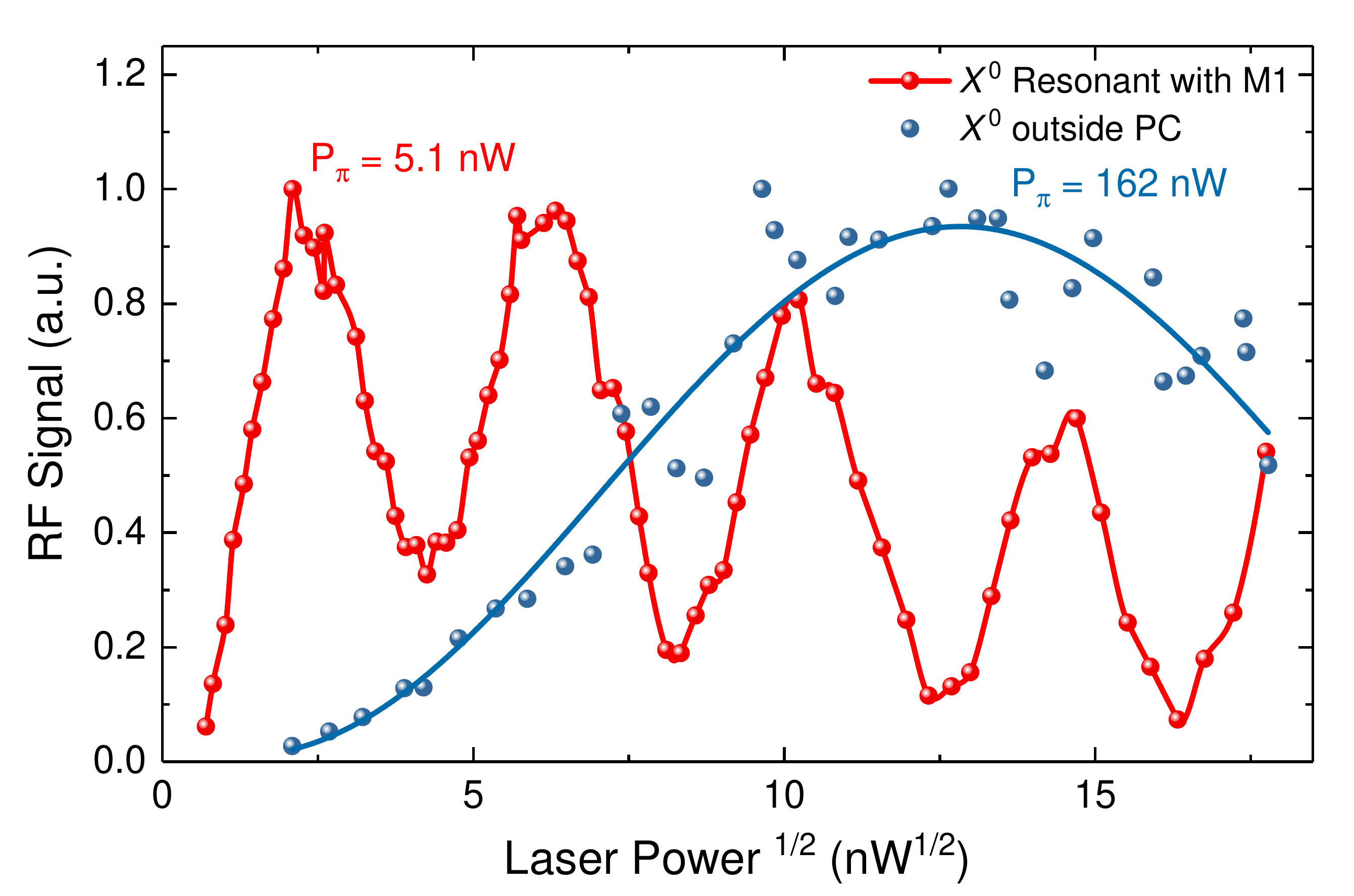}
\caption{Comparison between Rabi rotation data for the QD exciton resonant with the M1 cavity mode (red data) and a different QD exciton located outside the photonic crystal (blue data). It can be seen that the cavity acts to strongly enhance the electric field experienced by the QD, reducing the power required for a $\pi$-pulse by approximately a factor of 32.}
\label{fig:PiPower}
\end{figure}

Owing to the localized optical field enhancement, the cavity should also serve to strongly enhance the excitation efficiency by reducing the amount of laser power to reach population inversion (a $\pi$-pulse). To confirm this, we compare a Rabi rotation measured using the QD--cavity system studied in the main text to one measured on the neutral exciton of a different QD which is on the same sample but outside the photonic crystal. This is shown in Fig. \ref{fig:PiPower}. A decrease in $\pi$-pulse power of approximately 32 is found for the QD in the cavity, confirming this hypothesis. As expected, increasing $\pi$-power as a function of QD--cavity detuning was also observed when calibrating the pulse areas ($\Theta$) for detuned DPRF measurements. The resonant $\pi$-power of $5.1~\textrm{nW}$ (corresponding to a pulse energy of $67~\textrm{aJ}$) illustrates the low optical power requirements of the source compared to parametric down-conversion (PDC) sources, which typically are driven with mW powers.


\section{Resonant Rayleigh Scattering}
\label{sec:RRS}

Resonant Rayleigh scattering (RRS) refers to coherent scattering of single laser photons by a two-level system -- in this case the QD exciton (e.g. Matthiesen et al., Phys. Rev. Lett. 108, 093602, 2012). This section presents some additional details to support the data presented in Fig. 3 of the main text.

\subsection{Signal to Background and Emission Rate}
\label{subsec:rate}

\begin{figure}[h]
\includegraphics[width=0.6\textwidth]{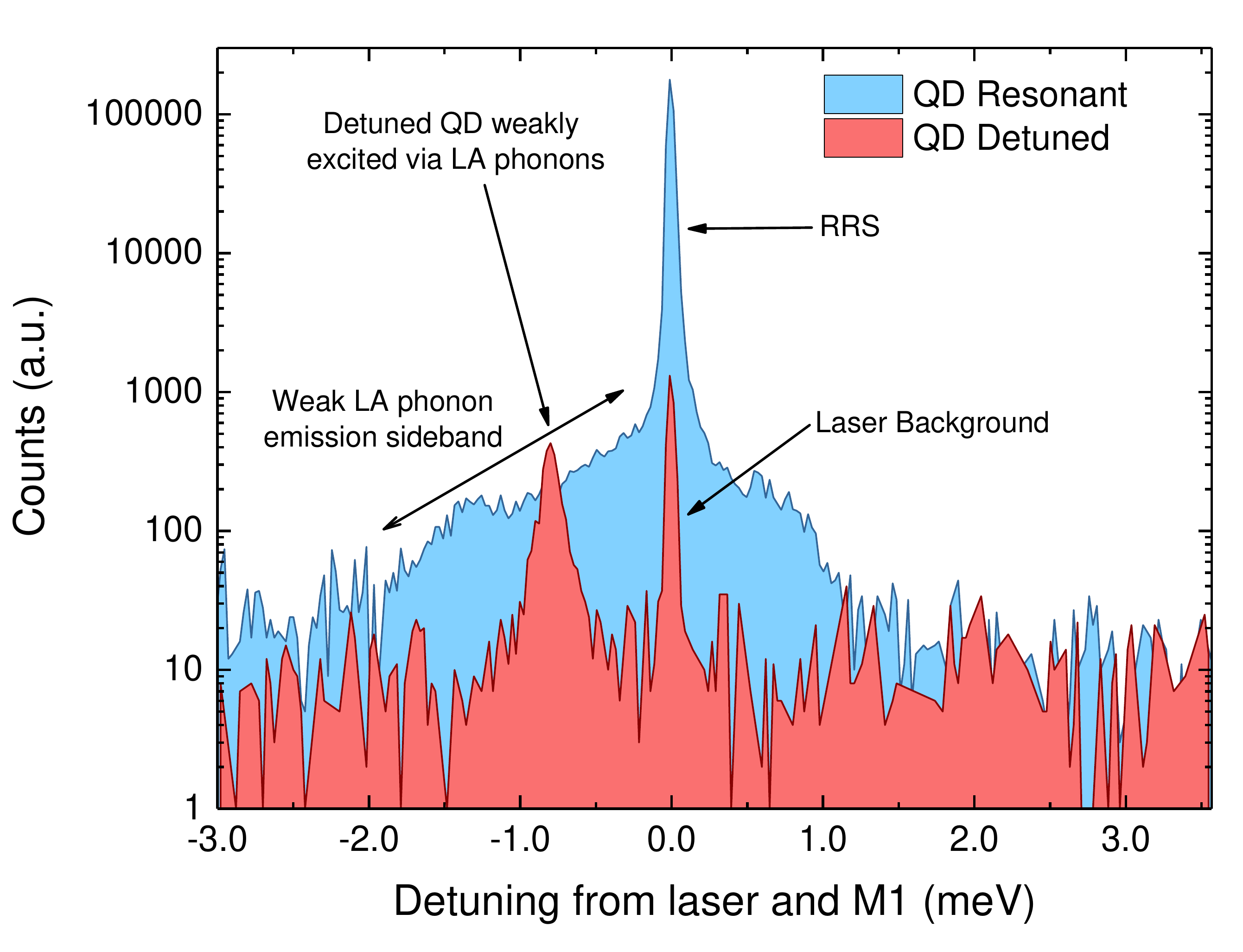}
\caption{Log-linear spectrum of the device under weak resonant CW excitation ($25~\mathrm{nW}$, $\Omega_{\textrm{R}}/2\pi \simeq 2~\textrm{GHz}$) when the QD is either resonant (blue data) or detuned (red data, detuning $-0.77$~meV) from the laser and M1 cavity mode. The cavity excitation / waveguide collection scheme used for the correlation measurements was also used here. As this spectrum was taken with a spectrometer and a CCD (as opposed to the FPI), it is not possible to resolve the RRS and RF components as they are both resolution-limited by the instrument.}
\label{fig:RRSBg}
\end{figure}

To determine the signal to background ratio in the RRS measurements, we compare spectra (taken with the spectrometer and CCD) with the QD resonant with and detuned from the laser, similar to the method shown for pulsed driving in Fig. 1(c) of the main text. The laser suppression is considerably stronger for the single mode CW laser as the narrow spectral width reduces the influence of birefringence in the optical setup. As a result, it is necessary to plot the intensity on a logarithmic scale for the laser background peak to be visible. This is shown for the case of cavity excitation and waveguide collection at a driving power of $25~\textrm{nW}$ ($\Omega_{\textrm{R}}/2\pi \simeq 2~\textrm{GHz}$) in Fig. \ref{fig:RRSBg}. In the Fabry--P\'erot measurements in the main text, an RRS fraction of $87.4~\%$ was found at this driving strength.

Comparison of the areas of the central peaks gives a signal to background ratio (SBR) of approximately 150:1. The absence of a significant peak at the detuning $\Delta=0$ (where $\Delta$ is the detuning relative to the laser and M1 cavity mode) in the QD detuned spectrum demonstrates the fundamental role that interaction between the emitter and laser plays in coherent scattering. When the QD is resonant, weak asymmetric sidebands corresponding to emission ($\Delta < 0$) or absorption ($\Delta > 0$) of a longitudinal acoustic (LA) phonon followed by spontaneous emission of a photon can be observed. It is also notable that in the detuned case a small amount of spontaneous emission from the zero-phonon line (ZPL) is still observed as the QD is weakly (owing to very small $\Omega_{\textrm{R}}$) excited via LA phonon emission (Quilter et al., Phys. Rev. Lett. 114, 137401, 2015).

In order to determine the count-rate in the waveguide in this regime where RRS is dominant, we measure the count-rate under the same conditions as the resonant data measured in Fig. \ref{fig:RRSBg}. To do this, a single SPAD is connected directly to the collection fiber, and a count-rate of $66.0 \pm 0.8~\textrm{kHz}$ is measured. Using FDTD simulations, the first lens is found to collect $14~\%$ of the light scattered by the out-coupler with $23~\%$ of this coupled into the single mode collection fiber. The beam splitter in the setup also causes a loss of $50~\%$ whilst the linear polarizer has a transmission of $84~\%$ for a perfectly co-polarized input. Finally, the SPAD has a quantum efficiency of $43~\%$ at the QD wavelength. Combining these losses, a collection efficiency of $0.58~\%$ is deduced, leading to an estimated waveguide count-rate of $11.5 \pm 0.4~\textrm{MHz}$ at this high RRS fraction.

\subsection{Analysis of the Fabry--P\'erot spectra}
\label{subsec:FP}

\begin{figure}[h]
\includegraphics[width=\textwidth]{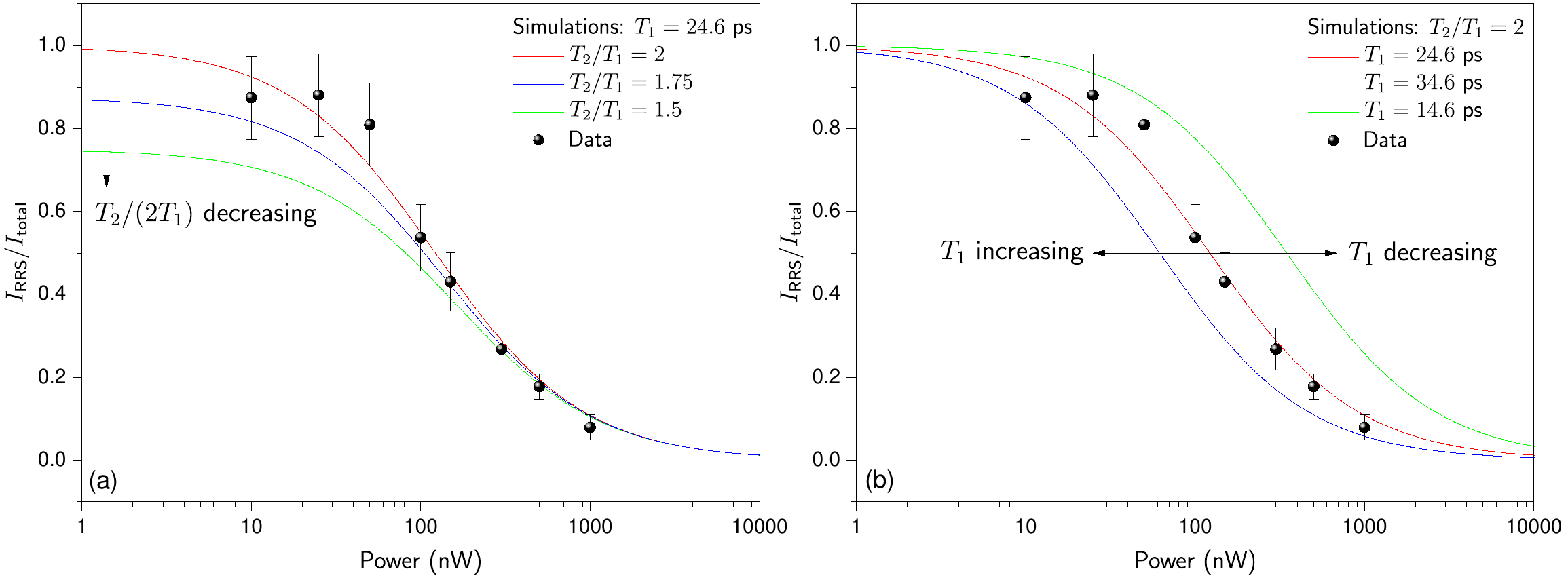}
\caption{Experimental $I_\text{RRS}/I_\text{total}$ (black spheres) and fits with Eq.~3 from the main text (coloured lines). (a) Illustration of the effect of reduced coherence. Here the curves have the same $T_1$ value, and together they show that a high fraction of coherent scatter at low power implies that the emitter coherence is very close to the radiative limit. For $T_2/T_1=1.5$ (green curve), for example, it is not possible to reach $80$~\% $I_\text{RRS}$. (b) Illustration of the effect of varying $T_1$. Comparing radiatively-limited curves ($T_2=2T_1$), we see that the point at which coherent scattering gives way to incoherent scattering is strongly dependent on $T_1$. This is a reflection of the fact that shorter lifetimes have higher saturation powers. For both (a) and (b) the red curve is the fit shown in Fig.~3 of the main text.}
\label{RRS2}
\end{figure}

This section provides further information on how the data for Fig. 3 was obtained. The Fabry--P\'erot spectra consisted of a series of peaks separated by the free spectral range (FSR). These have three components: RRS, SE, and laser background. At low power the laser background, observed by detuning the dot, is negligible ($0.5$~\% for $10$~nW). This background increases with power and is in all cases subtracted. A function consisting of the sum of a Lorentzian peak (for the SE) and a Gaussian peak (for the RRS) was fitted to the data. Here the Gaussian was used to approximate the Fabry--P\'erot instrument response function (IRF), from which the sub-IRF linewidth coherent scatter cannot be distinguished. At low powers the SE component is spectrally broad with negligible intensity, and the fits are therefore constrained using a linewidth obtained from higher power measurements. The 500 nW and 1000 nW SE components were adjusted to account for clipping of the signal as the Mollow side peaks approach the edge of the filtering window.

Figure 3 in the main text shows that the $I_\text{RRS}/I_\text{total}$ data is well reproduced by a fit of Equation 2 that results in values of $T_1=(24.6\pm1.6)~\text{ps}$ and $T_2/(2T_1) \sim 1$. Fig.~\ref{RRS2} shows that the theoretical curve is very sensitive to the values of both these quantities. The high fractions of RRS ($\sim 87~\%$) observed at low power are only possible if $T_2/(2T_1) \sim 1$ (Fig. \ref{RRS2}(a)), and $T_1$ determines the point at which incoherent scattering begins to dominate. With $T_1=14.6~\text{or}~34.6~\text{ps}$ this occurs much too late or early respectively (Fig. \ref{RRS2}(b)), showing the high sensitivity to $T_1$, and providing additional confirmation of the value of $T_1$ deduced from the DRPF measurements.

\subsection{Mollow triplet and Rabi frequencies}
\label{MtaRf}

\begin{figure}[h]
\includegraphics[width=\textwidth]{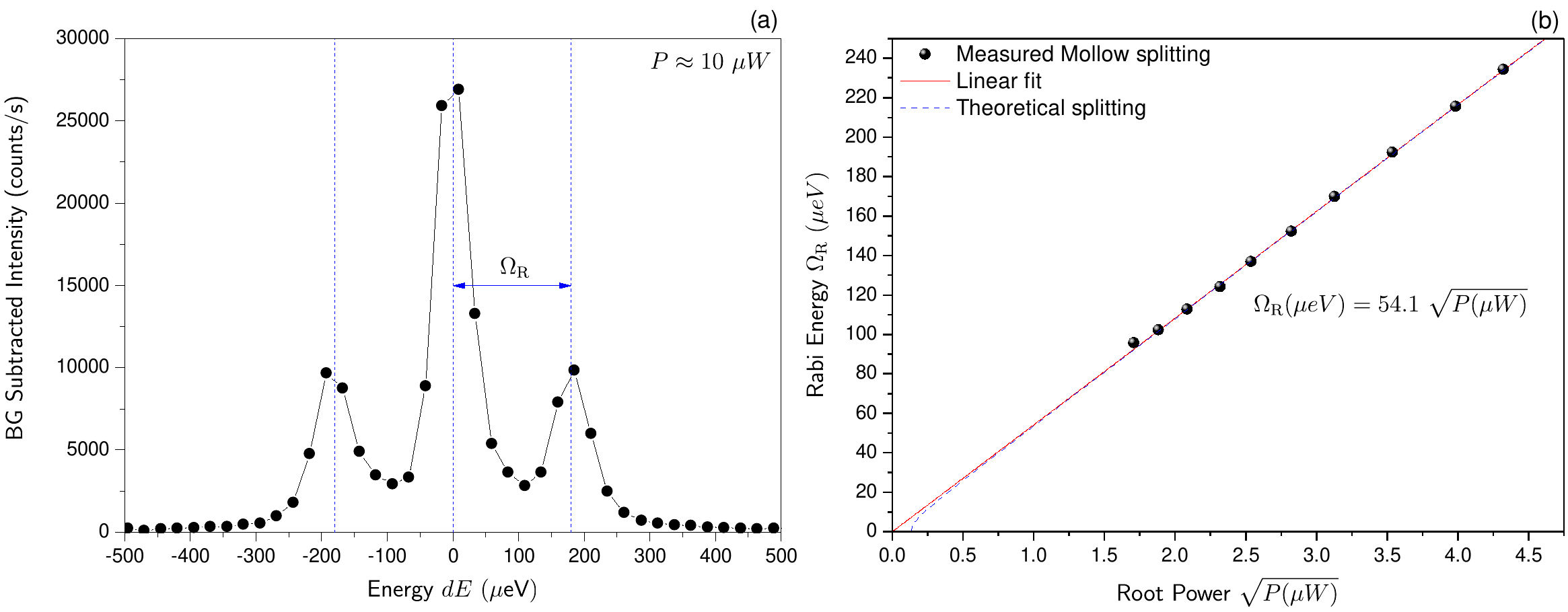}
\caption{(a) A high power ($\sim 10~\upmu$W) background-subtracted spectrum showing the very large Mollow splitting. (b) The measured dependence of the splitting on laser power and the deduced theoretical splitting from eq.~\ref{Eq: mollow}. At very low powers the splitting is damped and no triplet occurs.}
\label{mollow}
\end{figure}

As discussed in the main text, when $\Omega_\text{R} \ll \gamma_1$ we observe RRS. At high driving strengths the fraction of RRS reduces and eventually a Mollow triplet forms, as shown in Fig.~\ref{mollow}(a). This occurs when the damped Rabi frequency $\Omega_\text{R}^\text{d}$, given by (Muller et al., Phys. Rev. Lett. 99, 187402, 2007)

\begin{equation}
{\Omega_\text{R}^\text{d} = \sqrt{{\Omega_\text{R}}^2 - \dfrac{1}{4}\left(\gamma_1-\gamma_2\right)^2}},
\label{Eq: mollow}
\end{equation}

becomes real. The splitting is proportional to the square root of the power and allows us to extrapolate the Rabi frequencies down to the low powers of the RRS regime, as shown in Fig.~\ref{mollow}(b).


\section{Master equation simulations}
\label{sec:Master}

A Lindblad master equation (ME) for two-level system (2LS) cavity QED with coherent driving of the cavity mode is (Carmichael, Statistical Methods in Quantum Optics 2: Non-classical fields, Springer, 2008):

\begin{equation}
\begin{split}
\dot{\rho} = & - \dfrac{i}{2} \omega_A [\sigma_z,\rho] - i \omega_C[a^\dagger a, \rho] \\
& + g[a^\dagger \sigma_- - a\sigma_+,\rho] - i[\bar{E}_0e^{-i\omega_0 t}a^\dagger + \bar{E}_0^*e^{i\omega_0 t}a,\rho] \\
& + \dfrac{\gamma'_1}{2}(2\sigma_- \rho \sigma_+ - \sigma_+ \sigma_- \rho - \rho \sigma_+ \sigma_-) + \kappa(2a \rho a^\dagger - a^\dagger a \rho - \rho a^\dagger a),
\end{split}
\label{Eq: ME}
\end{equation}

\noindent where $g$, $\kappa$ and $\gamma'_1$ are defined in Section~\ref{dipole}, $\omega_A$ and $\omega_C$ are the angular frequencies of the 2LS and cavity respectively, and $\bar{E}_0$ and $\omega_0$ are the amplitude and frequency of the driving field. In this section it is used as a basis to:
\renewcommand{\theenumi}{\Alph{enumi}}
\begin{enumerate}[noitemsep]
 \item Compare the SPAD lifetime measurement to the DPRF and RRS measurements.
 \item Explain the discrepancy between resonant and non-resonant PL decay rates.
 \item Analyze the principle, the results, and the implications of the DPRF technique.
 \item Simulate the relationship between $\pi$-pulse duration and multiple emission events.
\end{enumerate}
The ME was solved and analyzed with the help of the Quantum Toolbox in Python (QuTiP) (Johansson et al., Comp. Phys. Comm. 184, 1234, 2013).
The pulses were modeled as Gaussians with electric-field temporal FWHMs $T_\text{P}$, and either excited the exciton directly, or through the cavity mode (as described by the Eq. \ref{Eq: ME}) -- for this system the difference between the exciton population dynamics is negligible, with the main effect being a difference in the cavity population (and required computational resource). Since the SBR was very good in the experiments, the background cavity population can be ignored to a good approximation. Additionally, because (as we will see) long $\pi$-pulses result in multiple emission events, in all cases a Rabi oscillation was simulated to determine the pulse area that gives the first maximum of emission -- i.e. the ``experimental $\pi$-power", which becomes increasingly larger than $\pi$ for increasing $T_\text{P}$. In subsections A, B and C, a $\pi$-pulse is that which gives the first simulated maximum of emission, even though the discrepancy for these pulse durations is small.

\subsection{SPAD lifetime measurement}
\label{subsec:lifetime}

\begin{figure}[h]
\includegraphics[width=0.6\textwidth]{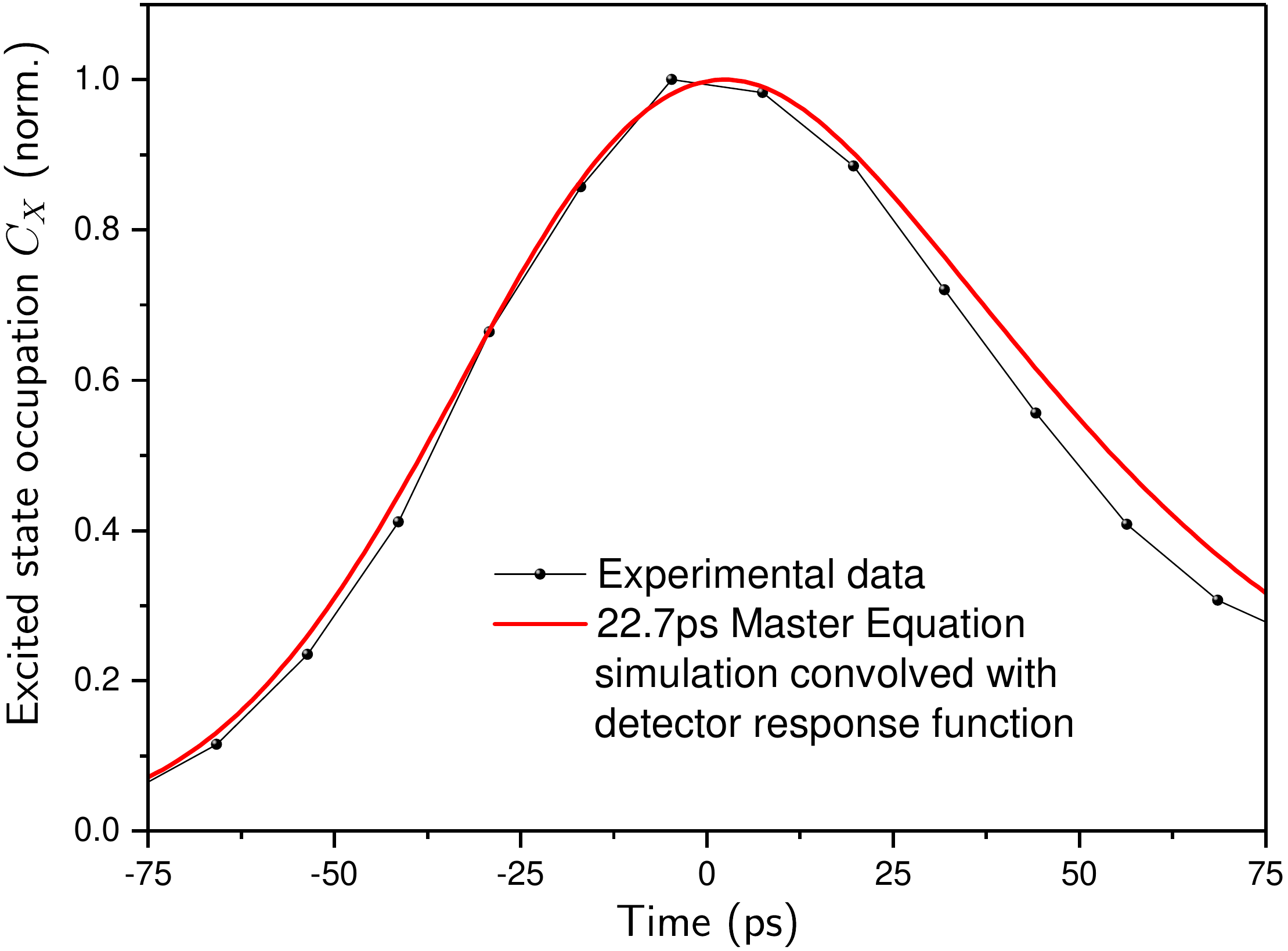}
\caption{Comparison of the lifetime measured with a SPAD (FWHM $\sim 60$~ps) to a simulation of the $22.7$~ps decay under $\pi$-pulse excitation, after convolving the simulation with a measured detector response function.}
\label{ME_SPAD}
\end{figure}

The SPAD lifetime measurements described in the main text revealed that the exciton lifetime was too short to reliably measure with the detector FWHM ($\sim 60$~ps). Nevertheless, once the lifetime was known via other techniques (DPRF and RRS), it was possible to simulate the pulsed population dynamics with the ME, convolve this with the IRF, and compare with the data. The results of this procedure are shown in Fig.~\ref{ME_SPAD}. The agreement is very good, and the small discrepancy is believed to be due to variabilities in the IRF, which changes with wavelength and spot size on the SPAD. These changes become significant when operating at or below the quoted limit of the detector. Nevertheless the SPAD measurements further justify the DPRF result and the lifetime extracted from the RRS.

\subsection{Comparison of resonant and non-resonant excitation decay dynamics}
\label{subsec:decay}

\begin{figure}[h]
\includegraphics[width=0.6\textwidth]{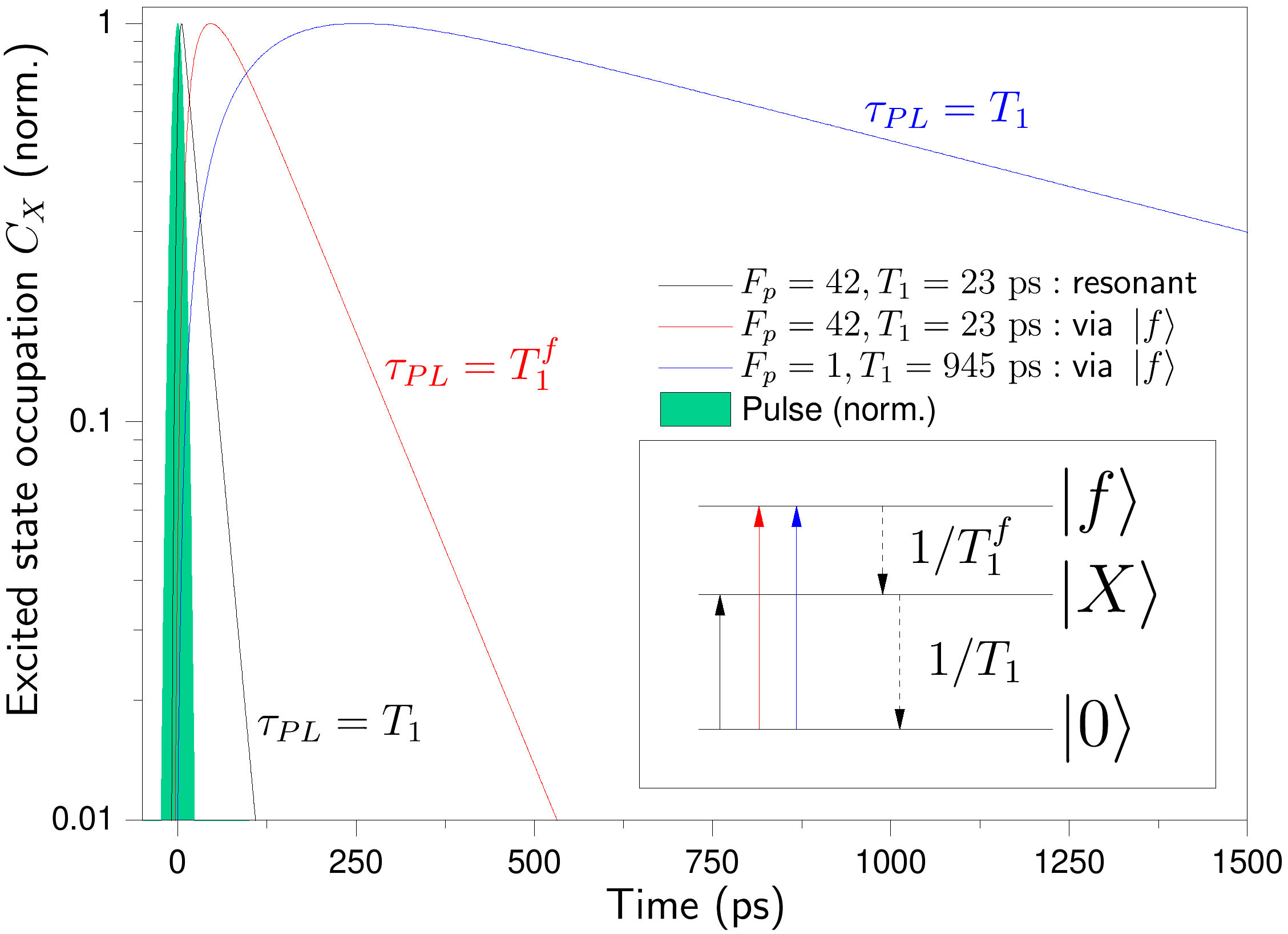}
\caption{$\left|X\right>$ population dynamics under various excitation conditions. When exciting resonantly (black curve), a fast rise and decay at the Purcell-enhanced rate (here $T_1=23$~ps with $F_p=42$) is observed. For excitation via a higher energy state $\left|f\right>$ which populates $\left|X\right>$ at a rate $1/T_1^f$, with $T_1^f=100$~ps (red curve), we see a slower rise and a decay rate of $100$~ps, i.e. the decay rate in this case is determined by the slow filling rate of the state. If we turn off the Purcell enhancement to make the $\left|X\right>$ decay time $945$~ps, and again fill the state via the now relatively fast decaying third higher level (blue curve), we see a very slow rise but what we measure at long times is again the true $\left|X\right>$ decay time of $945$~ps. Inset: Energy level diagram.}
\label{pshell}
\end{figure}

The effect on the time-resolved $\left|X\right>$ dynamics when exciting via a third higher energy state $\left|f\right>$ is shown in Fig.~\ref{pshell}. An additional collapse operator has been added to the ME to allow $\left|f\right>{\to}\left|X\right>$ decay at a rate $1/T_1^f$, where $T_1^f$ is the lifetime of the higher state. With resonant pulses (exciting $\left|0\right>{\to}\left|X\right>$ directly via the cavity mode), a fast rise and decay at the Purcell-enhanced rate is observed. When exciting $\left|X\right>$  via $\left|0\right>{\to}\left|f\right>$ with $T_1^f>T_1$, the observed decay rate of the $\left|X\right>$ population $\tau_{PL}$ is determined by the filling rate of the state, $1/T_1^f$, rather than the Purcell-enhanced decay rate $1/T_1$. For $T_1^f \ll T_1$, the time-resolved PL curve approaches the resonant case. Thus, a time-resolved pulsed PL measurement will determine the radiative transition rate and hence Purcell factor only when the radiative rate is the slowest process in the excitation-emission cycle. This explains the observed difference in the time-resolved PL decay observed under non-resonant and resonant excitation shown in Fig. 1(c), in the case of slow carrier relaxation.

\subsection{Double $\pi$-pulse resonance fluorescence}
\label{subsec:DPRF}

\begin{figure}[h]
\includegraphics[width=\textwidth]{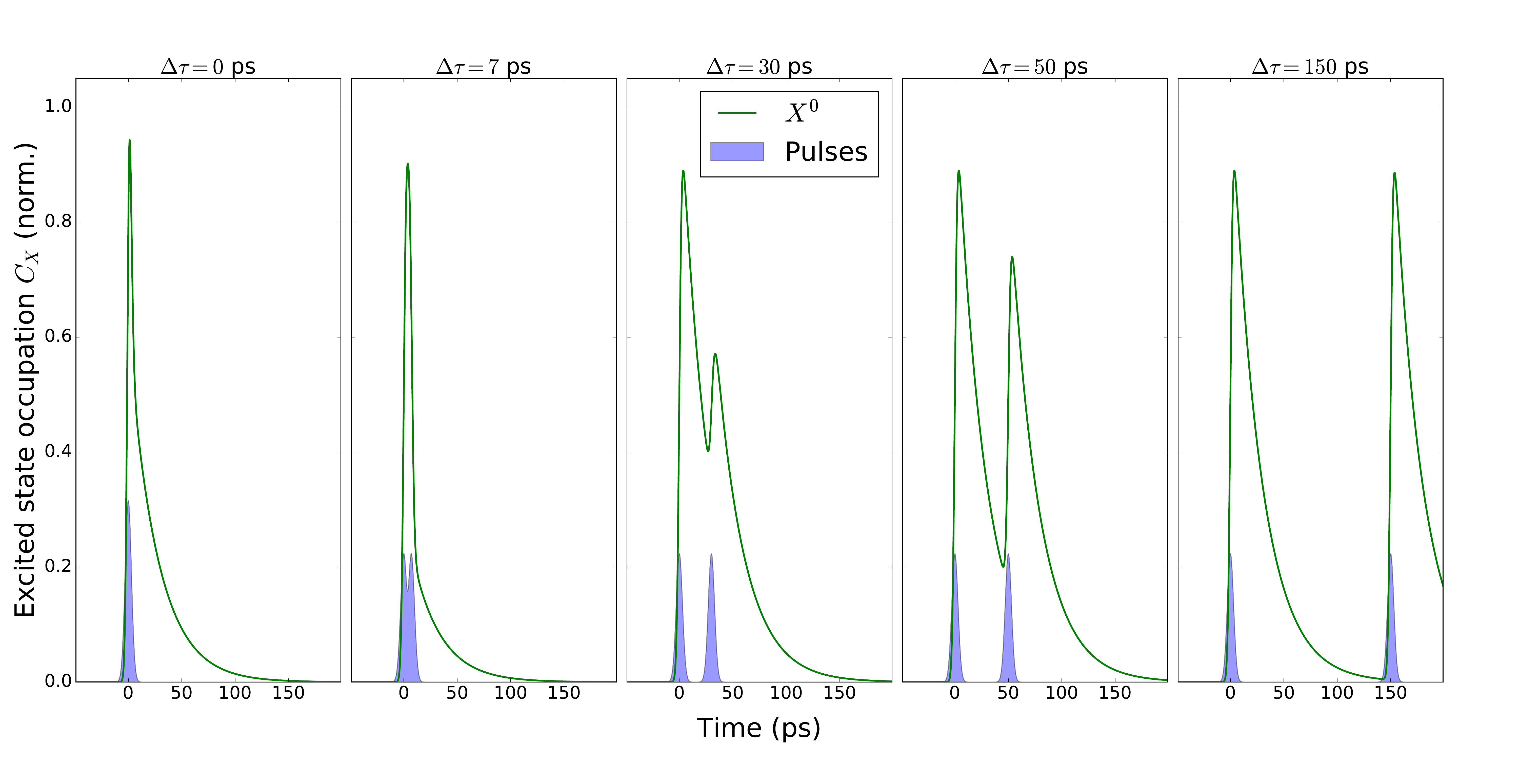}
\caption{The principle of the DPRF method shown via ME simulations of the system with two $7$~ps $\pi$-pulses. The total occupation probability is minimum around $\Delta t=7$~ps when the pulses just separate and can effectively populate and depopulate the state. The total population recovers exponentially with a time-constant given by the emitter lifetime.}
\label{Dpi_reel}
\end{figure}

\begin{figure}[h]
\includegraphics[width=0.7\textwidth]{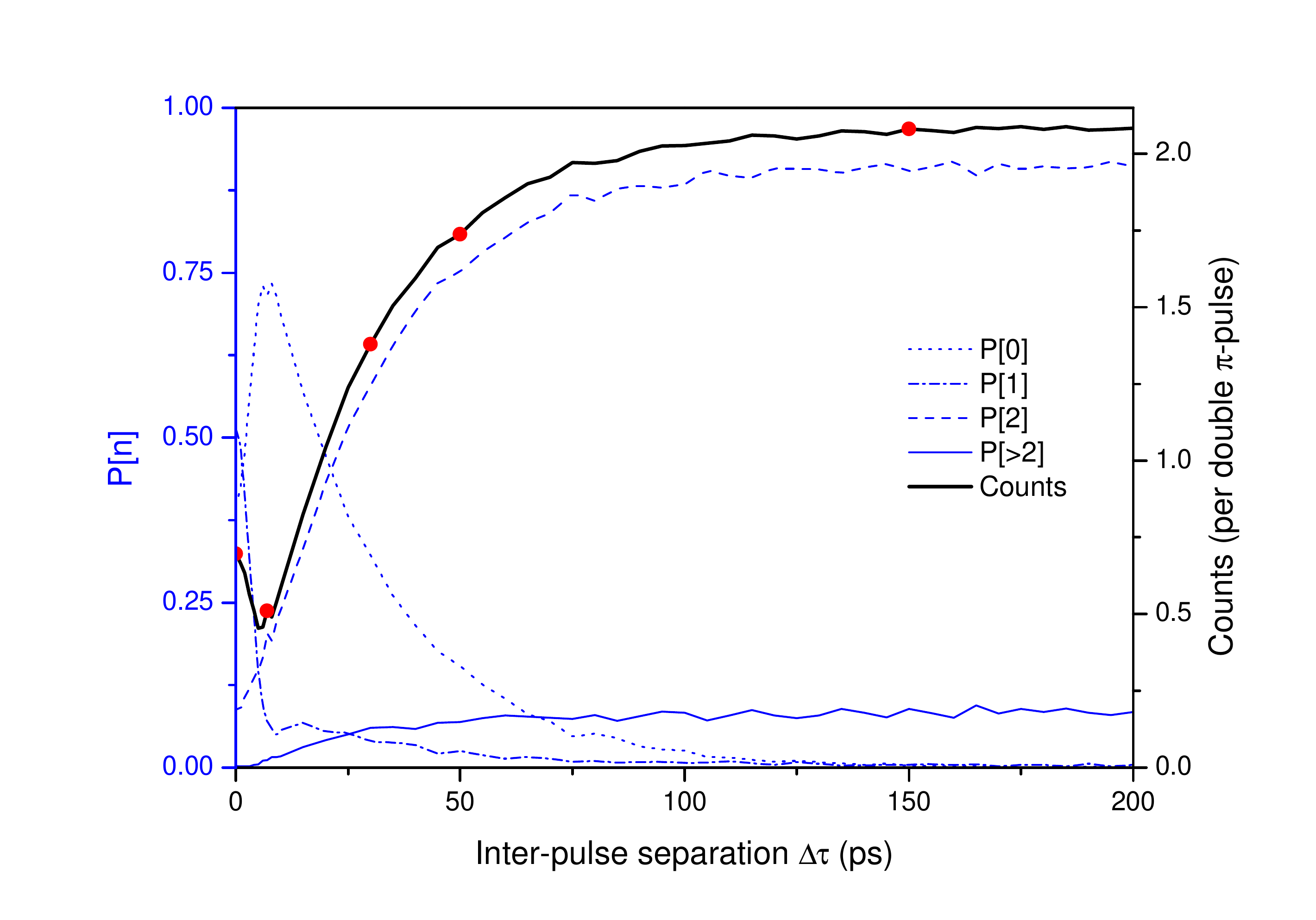}
\caption{Monte Carlo simulations of the DPRF technique. The black curve shows the expected counts, and the blue curves show the composition of the expected counts in terms of emission number probabilities. The simulations reveal that the expected signal recovers on the timescale of the emitter lifetime. The red points refer to the pulse separations depicted in Fig.~\ref{Dpi_reel}.}
\label{Dpi_int}
\end{figure}

The principle of the DPRF technique is illustrated in Fig.~\ref{Dpi_reel} via ME simulations of the time dynamics of the excited state $\left|X\right>$ for several inter-pulse separations. These separations are indicated by red dots in Fig.~\ref{Dpi_int}, which shows the expected counts and emission number probabilities.

The main features of DPRF are determined by the emitter time-constant $T_1=T'_1/F_\text{P}$, the pulse duration $T_\text{P}$, and the ratio of the two. The maximum instantaneous population inversion due to a single $\pi$-pulse is proportional to $T_1/T_\text{P}$. Thus the maximum depopulation due to the second $\pi$-pulse is also proportional to $T_1/T_\text{P}$, and the point at which this occurs is determined by $T_\text{P}$, since at $\Delta \tau=0$ the pulses combine to give a $\sqrt{2}\pi$-pulse. However, upon separation of the pulses, the recovery of the signal is determined only by $T_1$. As such, one can obtain the emitter lifetime even with $T_\text{P}>T_1$ provided one fits away from the region where the pulses overlap temporally. Experimentally, some additional noise may be seen around $\Delta t=0$ due to interference between the pulses as they are combined in the optical setup.

The solutions to the ME thus far have used the density matrix formalism and thus produced expectation values for ensemble averages. Now the Monte Carlo method is employed to gain insight into the quantum jumps that the system undergoes. In particular we are interested in the number of quantum jumps from the $\left|X\right>$ state to the ground state over the entire course of the two $\pi$-pulse system evolution for a single run of the system -- a single quantum trajectory. By counting the jumps of thousands of such trajectories we obtain a probability distribution for the number of quantum jumps, and therefore the number of emissions -- with some probability $P[0]$ we will get 0 photons after two $\pi$-pulses, some probability $P[1]$ we will get 1 photon etc. This is repeated for different inter-pulse separations. Fig.~\ref{Dpi_int} shows the emission number probabilities for different pulse separations (blue) and the average total number of photons per trajectory (black). Close to $\Delta\tau=0$, 0-emission trajectories dominate, and for $\Delta\tau > T_{1}$, 2-emission trajectories are the most probable. Except very close to $\Delta\tau=0$ (where the pulses interfere), 1-emission trajectories are very improbable -- showing that in general the $\pi$-pulses either both create a photon each or else cancel each other out. For the simulated pulse duration ($T_\text{P} \sim T_1 /4$) there is a small probability of multiple emission events for each $\pi$-pulse, and so the expected count is slightly larger than 2 for large pulse separations.

The double pulse simulations also highlight a point concerning emission number purity. As the dashed blue line in Fig.~\ref{Dpi_int} shows, the probability of two emissions increases with $\pi$-pulse separation on a time scale determined by the emitter lifetime. For negligibly short pulses
\begin{equation}
P[2] = 1 - e^{-\frac{\Delta\tau}{T_1}}.
\label{Eq: P2}
\end{equation}
For $\Delta\tau = 5 T_1$ the ``2-emission" purity is $99.3$~\%. By extension, very high emission number purity per pulse under N sequential $\pi$-pulses requires separations much longer than the emitter time constant. This therefore puts a stronger requirement on emitter lifetime for high $\pi$-pulse repetition rates.

\subsection{Relationship between $\pi$-pulse duration and $g^{(2)}_{HBT}(0)$}
\label{subsec:duration-HBT}

In the previous subsection it was seen that multiple emission events may occur for a single $\pi$-pulse. This is due to the possibility of having multiple excitation events over a finite pulse duration. As $T_\text{P}/T_1$ (the $\pi$-pulse duration relative to the excited state lifetime) increases, we expect that the probability of multiple excitations and hence multiple emissions increases. Assuming that the pulse duration is still less than the detector resolution, $g^{(2)}_{HBT}(0)$ will increase and the SPS will appear to have a non-ideal single-photon purity.

The relationship between $T_\text{P}/T_1$ and $g^{(2)}_{HBT}(0)$ was investigated through Monte Carlo trials of the system, as in the previous subsection. The emission number probabilities $P[n]$ can be used to calculate $g^{(2)}_{HBT}(0)$ through
\begin{equation}
g^{(2)}_\text{HBT}(\tau=0) = \dfrac{\sum_{0}^{\infty}n(n-1)P[n]}{\left(\sum_{0}^{\infty}nP[n]\right)^2}.
\end{equation}
Note that this formula is in direct correspondence to that for Fock state photon number distributions. This correspondence is valid because for the experimental case of low detector resolution relative to pulse duration, a 2-photon emission event is not distinguishable from two single-photon excitation-emission cycles within the pulse duration. The results of the simulations are shown in the inset to Fig. \ref{HOM}(a) of the manuscript. As expected, $g^{(2)}_{HBT}(0)$ increases with $T_\text{P}/T_1$, and excellent agreement is observed with the experimental HBT values.

In the simulations of Fig. \ref{HOM}(a), the $\pi$-pulse area was defined to be exactly $\pi$. As stated at the beginning of this section, the ``experimental $\pi$-power" becomes increasingly larger than $\pi$ for increasing pulse duration. Therefore, for very long pulses, the ``experimental $\pi$-power" $g^{(2)}_{HBT}(0)$ will increase even more rapidly than Fig. \ref{HOM}(a) suggests due to an even larger multiple excitation probability.


\section{Correlation Measurements}
\label{sec:correlation}

\begin{figure}[h]
\includegraphics[width=\textwidth]{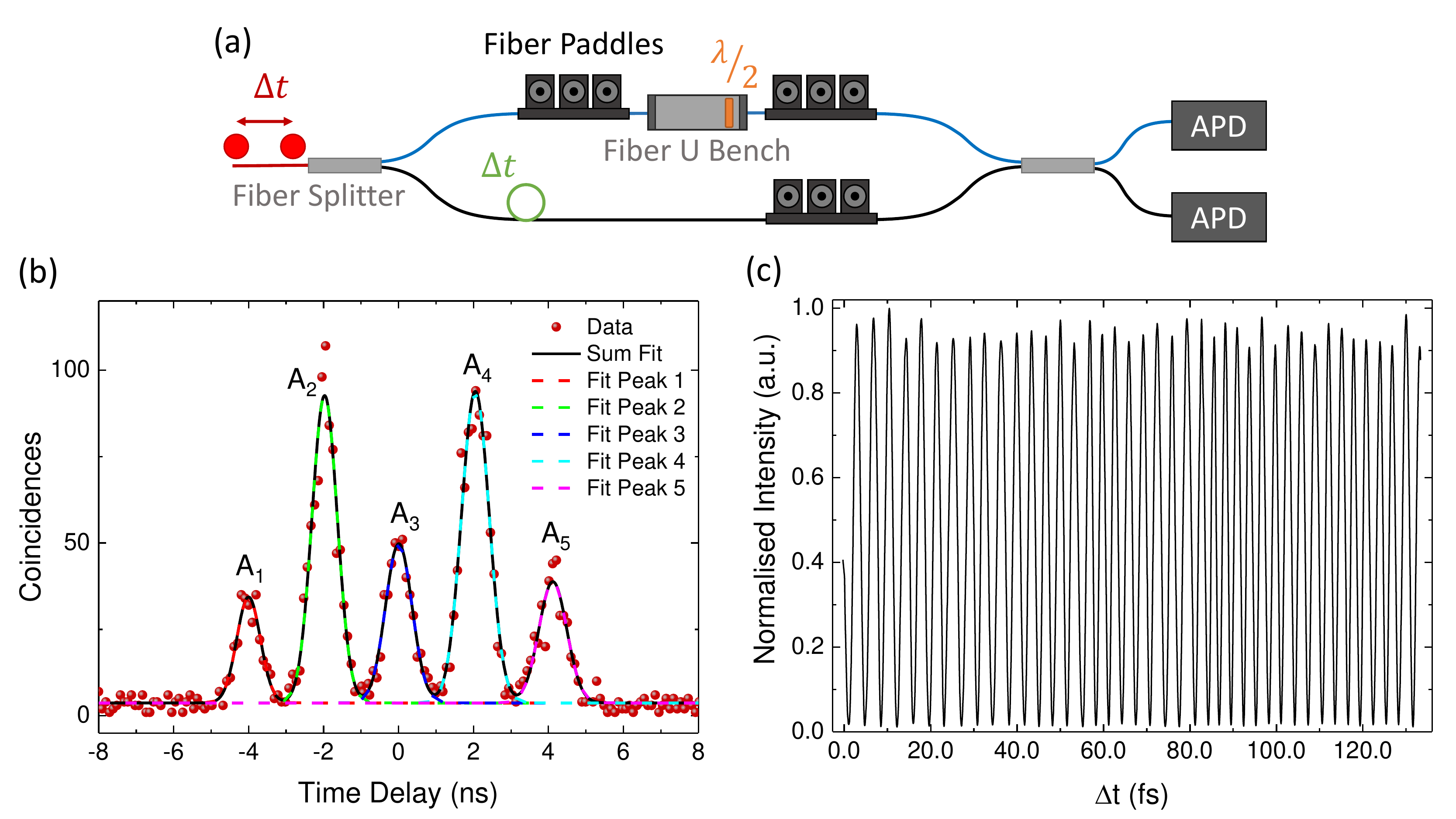}
\caption{(a) Schematic of the Mach--Zehnder interferometer used for the HOM measurements. (b) Coincidence count data for the co-polarized ($\parallel$) case. The dashed lines show the individual Gaussian (with width from the SPAD IRF) fits to each peak whilst the black line shows the cumulative fit. (c) Interference fringes measured by piezo tuning the path length of one interferometer arm. A single mode laser at the wavelength of the M1 mode is used and the interferometer is configured to be co-polarized with equal arm lengths. The transmission is monitored using a single detector on one output port of the interferometer.}
\label{fig:HOMSI}
\end{figure}

The Hanbury Brown and Twiss (HBT) and Hong-Ou-Mandel (HOM) measurements are performed using SPADs selected for maximum quantum efficiency ($\sim 43~\%$) at the cavity wavelength of $\sim 915~\textrm{nm}$. For the measurements with 13 ps pulses, the combined IRF of the two detectors when used with the photon counting card (TCSPCM) is Gaussian in shape with a FWHM of $\sim 860~\mathrm{ps}$. For the measurements with 2.4 ps pulses, this is $\sim 340~\mathrm{ps}$ owing to the use of newer detectors. The photon counting card is configured with a $50~\textrm{ns}$ delay window, corresponding to a time bin width of $48.9~\textrm{ps}$. A fixed electrical delay of $25~\textrm{ns}$ is added to one SPAD to centre the time-zero peak in the window.

For the HBT measurement the signal (either unfiltered for 13 ps pulses or filtered through the spectrometer for 2.4 ps pulses) is fed to a single fiber splitter with a SPAD on each output. For the HOM measurement a fiber interferometer is used in the Mach--Zehnder configuration as illustrated in Fig. \ref{fig:HOMSI}(a). Fiber paddles are used to correct for birefringence induced by the fibers, ensuring polarization matching at the second fiber splitter where photon coalescence occurs. A delay fiber is added to one arm to introduce a delay of $T_{HOM}$ with respect to the other. The delay time is chosen to be significantly larger than both the emitter lifetime and detector response time, ensuring well-resolved peaks. For $T_{HOM} >$ laser pulse separation (as is the case for the $T_{HOM} = 24~\text{ns}$ measurement), it is also necessary to carefully select $T_{HOM}$ such that peaks from adjacent cycles do not overlap with the zero time peak.  A motorized half-wave plate (HWP) allows the polarization of the other arm to be rotated between co- and cross-polarized with respect to the other, making the photons either maximally or minimally distinguishable. The waveplate is rotated between every 15 minute acquisition cycle to minimize the influence of any time-dependent drifts.

A characteristic series of 5 peaks is observed centered around zero time delay (Santori et al., Nature 419, 594-597, 2002) as shown in Fig. \ref{fig:HOMSI}(b). We denote the areas of these peaks as $A_n$, numbered from left to right (see Fig. \ref{fig:HOMSI}(b)). As the detector IRF is much greater than the QD lifetime ($22.7~\textrm{ps}$), the peaks can be well-fitted using Gaussian functions with the width of the detector response as shown in Fig. \ref{fig:HOMSI}(b). This contrasts to the typical case of small Purcell enhancement where the IRF and QD lifetime are similar and it is necessary to convolve the IRF with the exponential QD response. At zero delay on the TCSPCM, single photons from subsequent pulses interfere. Comparing the areas of this peak for the co- and cross-polarized cases allows extraction of the raw visibility according to eq. \ref{eq:RawVis}:
\begin{equation}
V = \frac{A_{3\bot} - A_{3\parallel}}{A_{3\bot}}.
\label{eq:RawVis}
\end{equation}

To extract the true visibility of the two-photon interference, it is necessary to correct for both the multi-photon emission of the source ($\textrm{g}^{\textrm{(2)}}(0)$) and deviations of the interferometer beam splitter from ideal behavior. The relevant parameters are $\textrm{g}^{\textrm{(2)}}_{HBT}(0)$, the interferometer fringe contrast $(1-\epsilon)$ and the beam splitter reflection and transmission coefficients ($R$, $T$). These parameters for our experiment are given in Table \ref{tab:VisParams}. The fringe contrast was measured by adding a piezo-tunable air-gap to one arm of the interferometer, equalizing the length and intensities of the two arms and measuring the transmission of a single mode laser (at the wavelength of the M1 mode) through the interferometer in the co-polarized configuration as a function of this delay. The raw data of this measurement is shown in Fig. \ref{fig:HOMSI}(c). The value in Table \ref{tab:VisParams} was obtained by finding the fringe contrast ($=(I_{max}-I_{min})/(I_{max}+I_{min})$) for each fringe and taking the mean.

\setlength{\tabcolsep}{0.5em}
\begin{table}[!h]
\begin{tabular}{| l | c | c | c |}
 \hline
 Parameter & Value & Correction & Measurement Method\\
 \hline
 $(1-\epsilon)$ & $0.968 \pm 0.004$ & $6.38~\%$ & Fringe contrast measurement with single mode laser\\
 \hline
 $\textrm{g}^{\textrm{(2)}}_{HBT}(0)$, $T_{P}=13~\textrm{ps}$ & $0.134 \pm 0.003$ & $19.5~\%$ & HBT measurement\\
 \hline 
  $\textrm{g}^{\textrm{(2)}}_{HBT}(0)$, $T_{P}=2.4~\textrm{ps}$ & $0.026 \pm 0.007$ & $4.0~\%$ & HBT measurement\\
 \hline 
 \multicolumn{1}{|l|}{$R$} & \multicolumn{1}{c|}{$0.544 \pm 0.002$} & \multirow{2}{*}{$1.53~\%$} & \multirow{2}{*}{Resonant transmission with single mode laser} \\\cline{1-2}
 \multicolumn{1}{|l|}{$T$} & \multicolumn{1}{l|}{$0.456 \pm 0.002$} & & \\
 \hline
 Polarisation & $99.99 \pm 0.01 \%$ & 0 & Resonant extinction with single mode laser\\
 \hline
\end{tabular}
\caption{Parameters used in the correction of the two-photon interference visibility. The contribution of each to the corrected value is estimated in the correction column. These values are approximate owing to the co-dependence of parameters in eqs. \ref{eq:SantoriVis} and \ref{eq:SomaschiVis}.}
\label{tab:VisParams}
\end{table}

The influence of these values is shown in eq. \ref{eq:SantoriVis} by their effect on the amplitude of the central peak in the HOM measurement (Santori et al., Nature 419, 594-597, 2002):

\begin{equation}
A_3 \propto (R^3T + RT^3)(1+2g^{(2)}(0))-2(1-\epsilon)^2R^2T^2V.
\label{eq:SantoriVis}
\end{equation}

By taking $V=1$ for $A_{3\parallel}$ and $V=0$ for $A_{3\bot}$ we can evaluate the raw visibility that would be measured for perfectly indistinguishable photons under these conditions. Our measured raw visibility can then be normalized by this to obtain the corrected value. Equivalently, it is also possible to perform the correction using a single formula that compares $A_{3\parallel}$ to $A_{2\parallel}$ and $A_{4\parallel}$ (eq. \ref{eq:SomaschiVis}) (Somaschi et al., Nat. Photon., 10, 340-345, 2016):

\begin{equation}
V = \frac{1}{(1-\epsilon)^2}\left[2g^{(2)}(0)+\frac{R^2+T^2}{2RT}-\frac{A_{3\parallel}}{A_{2\parallel}+A_{4\parallel}}\left(2+g^{(2)}(0)\frac{(R^2+T^2)}{RT}\right)\right].
\label{eq:SomaschiVis}
\end{equation}

Using the values from table \ref{tab:VisParams} and the unfiltered 13 ps HOM data, eq. \ref{eq:SantoriVis} yields a corrected visibility of $V=79.6\pm 5.9~\%$ whilst eq. \ref{eq:SomaschiVis} gives $V=79.8\pm 5.7~\%$. The dominant term in this correction is the non-unity purity of the emission characterised by $\textrm{g}^{\textrm{(2)}}_{HBT}(0)$, illustrating why there is a large improvement in the raw visibility by shortening the pulse duration. The presence of laser background is not corrected for in this approach (other than the contribution to $\textrm{g}^{\textrm{(2)}}_{HBT}(0)$); as such, these values represents a lower bound, limited by the scattered laser and uncertainty in the temporal overlap of the short photon wavepackets at the beamsplitter.

\newpage

\section{Estimated On-Chip Brightness}
\label{sec:brightness}

\setlength{\tabcolsep}{0.5em}
\begin{table}[!h]
\begin{tabular}{| l | c | c |}
 \hline
 Parameter & Value & Reference\\
 \hline
 Maximum Excitation Rep. Rate & 10~GHz & Fig. \ref{DPRF} and SI Section \ref{subsec:DPRF} \\
 \hline
  QD-Waveguide Coupling Efficiency & $40~\%$ & SI Section. \ref{sec:efficiency} \\
 \hline
 PhC Waveguide Propagation Loss (GaAs) & 17 dB/mm & Rigal et al. \textit{Optics Express} \textbf{25}, 28908 (2017) \\
 \hline 
  SNSPD Detection Efficiency (GaAs Best) & $20~\%$ & Sprengers et al. \textit{Applied Physics Letters} \textbf{99}, 10-13 (2011)\\
 \hline 
 SNSPD Detection Efficiency (Si Best) & $91~\%$ & Pernice et al. \textit{Nature Communications} \textbf{3}, 1325 (2012)\\
 \hline
\end{tabular}
\caption{Values and references for parameters used to estimate the potential on-chip count rate of the SPS.}
\label{tab:OnChipBrightness}
\end{table}

This section contains values and references for the parameters that are used to estimate the potential on-chip brightness in the Discussion section. To arrive at the figures in the main text of 4~MHz for 76.2 MHz excitation and 540~MHz for 10~GHz excitation, the SNSPD detection efficiency is conservatively taken to be $20~\%$, the best currently demonstrated for GaAs waveguides. As can be seen in the table, this value can reach $91~\%$ for Silicon waveguide devices which would increase both values by a further factor of 4.6. This is a reflection of the maturity of the Silicon platform rather than any intrinsic limitation of GaAs-based devices.

\end{document}